\documentclass[aps,pra,reprint,superscriptaddress,nofootinbib,longbibliography]{revtex4-2}
\usepackage{amsmath,amssymb,amsfonts,bm}
\usepackage{graphicx}
\usepackage{hyperref}
\hypersetup{hidelinks,pdfborder={0 0 0}}
\usepackage{physics}
\usepackage{braket}
\usepackage{dsfont}

\begin{document}

\title{From local weight selection to Zeno slowdown in an open Su-Schrieffer-Heeger chain with a single local loss}
\author{Y. T. Wang}
\affiliation{College of Physics and Materials Science, Tianjin Normal University, Tianjin
300387, China}
\author{X. Z. Zhang}
\email{zhangxz@tjnu.edu.cn}
\affiliation{College of Physics and Materials Science, Tianjin Normal University, Tianjin
300387, China}
\affiliation{Interdisciplinary Center, Tianjin Normal University, Tianjin 300387, China}

\begin{abstract}  We study a quadratic open SSH chain with a single-site loss and show that the many-body fermionic Lindblad problem admits an exact reduction to a finite non-Hermitian one-body matrix with a rank-one imaginary impurity. Its rapidities generate the complete Liouvillian spectrum and reveal three mechanisms governing the slowest relaxation. At weak loss, decay is selected by the clean local spectral weight at the lossy site, yielding the generic law $\Delta_{\mathcal L}\sim\gamma N^{-3}$ and, in the topological regime, exponentially smaller edge-controlled gaps. At intermediate loss, a centered bulk-loss geometry reaches an exact exceptional point on the real-$\gamma$ axis. Symmetry-related rapidity pairs coalesce simultaneously, including the pair at the lower rapidity edge. Exact real-space dynamics at this lower-edge exceptional point exhibits a polynomially enhanced exponential tail, whereas a matched high-energy control with the same parity-even one-body decay edge but no lower-edge defectiveness remains nearly exponential. At strong loss, one ultrafast defect mode separates from an active slow sector governed by a cut-chain Zeno problem, giving $\Delta_{\mathcal L}\sim\gamma^{-1}$ up to a geometry-dependent prefactor. The full finite fermionic Liouvillian spectrum, including its operator-parity sectors and subset-sum structure, is statistics-specific. By contrast, the elementary one-body decay spectrum and the three associated mechanisms are governed by a finite-dimensional linear drift matrix, so their spectral and dynamical signatures can also be accessed in bosonic and classical-wave platforms engineered to realize the same effective matrix. These results establish how topology, defect geometry, and local dissipation jointly organize long-time relaxation in an open dimerized lattice.  \end{abstract}
\maketitle
\section{Introduction}

Open quantum systems have moved well beyond the viewpoint that dissipation is
merely a nuisance. In the Lindblad framework, engineered coupling to the
environment can prepare states, stabilize phases, and reorganize dynamical
spectra in ways that have no closed-system analogue~\cite{Lindblad1976,Diehl2008,Verstraete2009}. In parallel, programmable superconducting circuits,
photonic lattices, and ultracold-atom platforms now provide direct control
over lattice geometry, local couplings, and site-selective loss
channels~\cite{Blais2021,Ozawa2019,Cooper2019}. This makes local dissipation a
practical control knob for nonequilibrium quantum simulation~\cite{shibata2019,shibata2020,guo2018,nakagawa2021,vznidarivc2015,zhang2020,tarantelli2021,landi2022,zhou2021,zheng2023,peng2024,zheng2025}.

For relaxation, the key spectral quantity is the Liouvillian gap. It is the smallest nonzero decay rate of the full generator and therefore sets the longest intrinsic relaxation time. A particular initial state or observable displays this rate only when it has nonzero overlap with the corresponding Liouvillian eigenmode; conserved operator-parity sectors can otherwise expose a faster sector-resolved decay scale~\cite{zhou2022,mori2023,sa2020}. More generally, Liouvillian spectral theory ties slow relaxation,
metastability, and dissipative critical behavior to the low-lying structure
of the Lindblad generator~\cite{Minganti2018,bao2026}. Direct time evolution can of
course extract the slowest scale, but by itself it does not reveal which
modes are selected by a local defect or how the slow sector is reorganized as
the loss is varied. What is needed instead is a formulation in which the
long-time dynamics can be traced to an explicit spectral problem. 

The present model admits such a formulation exactly. We consider a quadratic
open SSH chain with a single bulk loss~\cite{SSH1979,li2014,meier2016}. Because the Hamiltonian is quadratic
and the jump operator is linear, the many-body Lindblad problem can be
reduced by third quantization to a finite non-Hermitian single-particle
matrix with a rank-one imaginary impurity~\cite{Prosen2008,Prosen2010,prosen2011,guo2017,yamanaka2023}. Its
rapidities generate the full Liouvillian spectrum, and the lower rapidity
edge determines the slowest many-body decay rate. In this sense, the
relaxation problem of the original many-body system becomes an exact impurity
spectral problem rather than an effective approximation. This reduction lets us separate three regimes in a unified way. In the
weak-loss regime, decay is selected by the clean local spectral weight at the
lossy site. The generic result is $\Delta_{\mathcal L}\sim \gamma N^{-3}$,
whereas in the topological phase the gap becomes exponentially smaller when
the loss overlaps only exponentially weakly with an edge mode. In the intermediate regime, the exact secular equation shows that an exceptional point on the real-$\gamma$ axis requires balanced local spectral weights. In the centered-loss geometry studied here, chiral symmetry enforces this balance for symmetry-related pairs, and the spectrum reaches an exact exceptional point at $\gamma_{\mathrm{EP}}\simeq2.6200$. Among the simultaneous coalescences, the pair at the lower rapidity edge controls the gap and produces the defective late-time prefactor. A matched high-energy spectral control with the same parity-even decay edge but no lower-edge coalescence isolates this effect. In the strong-loss regime, one
ultrafast defect mode separates from the rest, while the remaining slow
sector is governed by a geometry-dependent Zeno problem on the chain cut by
the lossy site. Here the word Zeno does not mean that the entire chain becomes frozen. The local-loss channel instead creates a fast subspace: an amplitude entering site $s$ decays on the time scale $\tau_s=2/\gamma$. When $\gamma$ is the largest local scale, this rapidly emptied state is adiabatically eliminated from the slow dynamics. Propagation through it survives only as a virtual process of order $J_LJ_R/\gamma$, where $J_L$ and $J_R$ are the hoppings to its two neighbors. The lossy site is therefore projected out at leading order, the chain is effectively cut, and both the residual cross-defect transfer and the induced slow decay decrease as $1/\gamma$. This environment-induced projection of a rapidly decaying local subspace is the dissipative quantum-Zeno mechanism used in this work~\cite{Misra1977,FacchiPascazio2002,Syassen2008,Barontini2013}; the rest of the chain remains dynamically active.

Our aim is therefore not to treat the local loss as a small perturbation or
to emphasize exceptional-point phenomenology by itself. Rather, we show how a
single tunable defect organizes the full relaxation problem through three
mechanisms: local-weight selection at weak loss, lower-edge rapidity
reorganization at intermediate loss, and cut-chain Zeno reduction at strong
loss. This sequence identifies how topology, geometry, and dissipation act
together to determine the slowest relaxation scale in an exactly solvable
open SSH chain. It also places this problem at the intersection of
reservoir-engineered quantum matter~\cite{Diehl2011}, quadratic open-system
spectral theory~\cite{Prosen2008,Prosen2010}, and non-Hermitian topological
spectral reorganization~\cite{Bergholtz2021}.

We finally test which parts of this three-regime structure survive chiral-symmetry-preserving bond disorder. The weak-loss local-weight rule and the strong-loss $\gamma^{-1}$ Zeno law remain unchanged at the level of their asymptotic exponents. Within the finite chains and disorder range sampled here, exact imaginary-axis exceptional points also survive realization by realization, but their positions become sample dependent and their lower-edge dominance is progressively reduced. This distinction between robust asymptotic mechanisms and a disorder-sensitive intermediate spectral ordering is quantified in Appendix~\ref{app:disorder}.

The remainder of this paper is organized as follows. Section II introduces the
model and states the exact reduction to the finite rapidity problem. Section III
derives the scalar rapidity equation, and Sec.~IV discusses impurity-induced
bound rapidities and their eigenvectors. Sections V, VI, and VII analyze the
weak-, intermediate-, and strong-loss regimes, respectively. Section VIII
concludes the paper. Technical steps of the third-quantization construction and
the exact one-body dynamics used for the real-time simulations are collected in
Appendices A and B, respectively, while Appendix~\ref{app:disorder} examines
the robustness of the three-regime picture against weak bond disorder.

\section{Model, Liouvillian structure, and exact reduction}

\label{sec:model}

\subsection{Quantum simulation setting and effective open SSH model}

\label{subsec:model_H}

We consider a minimal quantum simulation architecture for dissipative transport and relaxation in a dimerized one-dimensional lattice. The setup consists of a programmable array with alternating nearest-neighbor couplings and a single site-resolved dissipative channel attached to a bulk site. Depending on the physical implementation, the lattice modes may represent resonators in a superconducting or photonic array, sites of a cold atom superlattice, or modes in a synthetic dimension, while the local loss can be generated by reservoir engineering, auxiliary lossy modes, or controlled particle removal. At the level of linear amplitudes and one-body correlations, these implementations share the same non-Hermitian drift matrix. The complete finite-dimensional operator-parity and subset-sum construction developed below, however, refers specifically to the fermionic Fock-space Lindblad problem.

The effective lattice contains $N$ unit cells with two sublattice sites $A$ and $B$ in each cell. The operators $c_{n,\alpha}$ and $c_{n,\alpha}^{\dagger}$ annihilate and create a spinless fermion on cell $n=1,\dots,N$ and sublattice $\alpha\in\{A,B\}$, and satisfy the canonical anticommutation relations. The Hamiltonian is
\begin{equation}
\begin{aligned}
H={}&\sum_{n=1}^{N} t_1 \left(c_{n,A}^{\dagger}c_{n,B}+\mathrm{h.c.}\right)\\
&+\sum_{n=1}^{N-1} t_2 \left(c_{n+1,A}^{\dagger}c_{n,B}+\mathrm{h.c.}\right),
\end{aligned}
\label{eq:H_SSH}
\end{equation}
where $t_1$ and $t_2$ denote the intracell and intercell hopping amplitudes. Open boundary conditions are imposed by the finite range of the second sum.

For later use, we map the two sublattices onto a single site index $j=1,\dots,2N$ through
\begin{equation}
c_{2n-1}\equiv c_{n,A},\qquad
c_{2n}\equiv c_{n,B},
\qquad n=1,\dots,N.
\label{eq:site_index}
\end{equation}
Defining the column vector $\bm c=(c_1,c_2,\dots,c_{2N})^T$, the Hamiltonian can be written as
\begin{equation}
H=\bm c^{\dagger} h\, \bm c,
\label{eq:H_single_particle}
\end{equation}
where $h$ is a $2N\times 2N$ Hermitian hopping matrix with nonzero elements
$h_{2n-1,2n}=h_{2n,2n-1}=t_1$ for $n=1,\dots,N$ and
$h_{2n,2n+1}=h_{2n+1,2n}=t_2$ for $n=1,\dots,N-1$. Dissipation is introduced through a single local jump operator placed on a bulk $A$ site. For a loss channel acting on the $A$ site of cell $m$, with $1<m<N$, we write
\begin{equation}
L=\sqrt{\gamma}\,c_{m,A}=\sqrt{\gamma}\,c_s,
\qquad
s\equiv 2m-1,
\qquad
\gamma\ge 0.
\label{eq:jump_bulk}
\end{equation}
The density matrix then evolves according to the Lindblad master equation
\begin{equation}
\partial_t \rho=\mathcal L[\rho]
=-i[H,\rho]+L\rho L^{\dagger}-\frac{1}{2}\{L^{\dagger}L,\rho\}.
\label{eq:lindblad_master}
\end{equation}

This formulation makes the physical content of the model transparent. The alternating hoppings encode the dimerized lattice geometry, while the local reservoir selects one bulk site as a tunable dissipative defect. The resulting competition between coherent hopping and local particle loss is the basic mechanism studied throughout this work.

Because $H$ is quadratic and $L$ is linear in the fermion operators, the Liouvillian remains quadratic and can be solved exactly by third quantization. Throughout the spectral analysis we use the standard full-operator-space Liouvillian gap
\begin{equation}
\Delta_{\mathcal L}\equiv \min_{\Lambda\in\operatorname{spec}(\mathcal L),\,\Re\Lambda<0}\bigl[-\Re \Lambda\bigr],
\label{eq:Liouvillian_gap_def}
\end{equation}
which is the smallest positive decay rate of the complete Liouvillian spectrum. The Liouvillian preserves operator parity. Consequently, a parity-even density matrix and a parity-even observable evolve inside the even operator sector and may be governed by the corresponding sector-resolved decay edge rather than by a parity-odd mode that realizes the full-space gap. The physical Lindblad loss rate is the parameter $\gamma$ in $L=\sqrt{\gamma}\,c_s$. A conditional one-particle amplitude on the lossy site decays at rate $\gamma/2$, whereas its population decays at rate $\gamma$. In third quantization the elementary rapidity blocks are $M=-ih-\Gamma/2$ and $M^*=ih^*-\Gamma/2$, with $\Gamma=\gamma\Pi_s$. The plotted matrix $P=h+i(\gamma/2)\Pi_s$ is one sign-conjugated rapidity block, not the no-jump Hamiltonian itself. Appendix~\ref{app:third_quantization} derives both parity sectors explicitly and shows how their union generates the complete $4^{N_s}$-dimensional Liouvillian spectrum, where $N_s=2N$.

Figure~1 summarizes this logic in two compact steps. Panel~(a) combines the quantum simulation architecture with the effective open SSH model. The dimerized chain, the singled-out bulk $A$ site $m$, and the local reservoir coupling $L=\sqrt{\gamma}\,c_{m,A}$ already show that the physical platform and the theoretical model are the same quadratic problem written at two complementary levels. Panel~(b) shows the exact reduction used throughout the paper: the many-body Lindblad equation is mapped exactly, by third quantization, to two $N_s$-dimensional rapidity blocks, represented by $P_\pm=\pm h+i(\gamma/2)\Pi_s$. Chiral symmetry makes $P_+$ and $P_-$ isospectral, so the single plotted matrix $P\equiv P_+=P_0+i(\gamma/2)\Pi_s$ contains all distinct rapidity values. Each eigenvalue $E_j$ of $P$ occurs once in each block, yielding the complete $2N_s$-rapidity multiset $\beta_j^{(\pm)}=iE_j$. Fermionic subset sums of these $2N_s$ elementary exponents generate the full $4^{N_s}$-dimensional Liouvillian spectrum. The factor $\gamma/2$ is the amplitude-loss rate produced by the jump operator $L=\sqrt{\gamma}\,c_s$, whereas the population on the lossy site decays at the physical rate $\gamma$. The smallest positive lower-edge value fixes the full-operator-space gap, $\Delta_{\mathcal L}=\min_{j:\,\operatorname{Im}E_j>0}\operatorname{Im}E_j$.

\begin{figure}[t]
\centering
\includegraphics[width=\columnwidth]{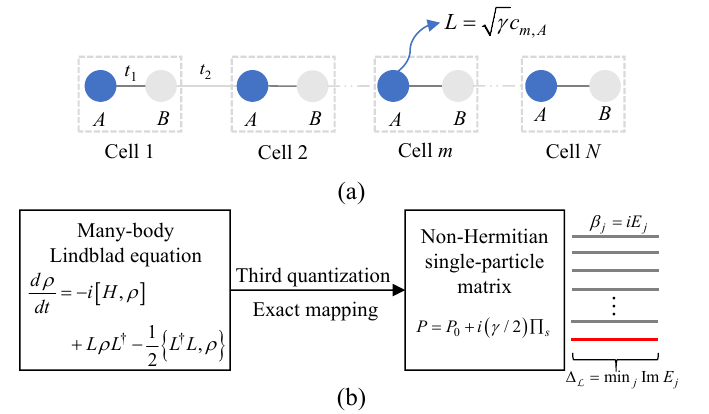}%
\caption{Two-panel summary of the model and the exact spectral reduction. (a) Quantum simulation architecture and effective open SSH model. A finite dimerized chain with intracell hopping $t_1$ and intercell hopping $t_2$ is coupled to a single local reservoir on the bulk $A$ site $m$. The dissipative defect is described by the jump operator $L=\sqrt{\gamma}\,c_{m,A}$. (b) Exact reduction schematic. Third quantization produces two rapidity blocks $P_\pm=\pm h+i(\gamma/2)\Pi_s$. Because chiral symmetry makes them isospectral, the plotted matrix $P\equiv P_+=P_0+i(\gamma/2)\Pi_s$ contains all distinct eigenvalues $E_j$. Each $E_j$ occurs in both blocks, so the complete rapidity multiset is $\beta_j^{(\pm)}=iE_j$. Fermionic subset sums of these $2N_s$ rapidities generate all $4^{N_s}$ Liouvillian eigenvalues, and the full-operator-space gap is $\Delta_{\mathcal L}=\min_{j:\,\operatorname{Im}E_j>0}\operatorname{Im}E_j$.}
\label{fig:framework}
\end{figure}

\subsection{Exact reduction to a finite non-Hermitian matrix}

\label{subsec:exact_reduction_main}

Because $H$ is quadratic and $L$ is linear in the fermion operators, the Liouvillian can be solved exactly by third quantization. The technical derivation is deferred to Appendix~\ref{app:third_quantization}; for the main text, it is sufficient to state the resulting finite-dimensional problem.

The elementary spectral problem of the many-body Lindblad generator reduces
exactly to two finite non-Hermitian one-body rapidity blocks. For the present
SSH chain these two blocks are isospectral, so all distinct rapidity values can
be obtained from a single $N_s\times N_s$ matrix acting in the spatial basis
$\bm\psi=(\psi_{1,A},\psi_{1,B},\dots,\psi_{N,A},\psi_{N,B})^T$; the complete
many-body Liouvillian spectrum is then reconstructed from fermionic subset sums
of the two-block rapidity multiset. We choose the plotted block
\begin{equation}
P(t_1,t_2;\gamma,m)=P_0(t_1,t_2)+\frac{i\gamma}{2}\,\Pi_s,
\qquad
\label{eq:P_imp_def}
\end{equation}
where $P_0$ is the clean SSH hopping matrix, $\Pi_s\equiv \bm e_s\bm e_s^T$ and $\bm e_s$ is the standard basis vector on the impurity site $s=2m-1$. Its only nonzero entries are
\begin{equation}
\begin{split}
(P_0)_{2n-1,2n}=(P_0)_{2n,2n-1}=t_1,\\
(P_0)_{2n,2n+1}=(P_0)_{2n+1,2n}=t_2,
\end{split}
\label{eq:P0_elements}
\end{equation}
for $n=1,\dots,N$ and $n=1,\dots,N-1$, respectively. The loss enters this spectral matrix through the rank-one diagonal term $i(\gamma/2)\Pi_s$. This matrix is not the no-jump Hamiltonian $h_{\mathrm{eff}}=h-i(\gamma/2)\Pi_s$ itself. It is the sign-conjugated block used in the rapidity convention below, chosen so that decay rates are positive numbers $\operatorname{Im}E_j$.

Let $E_j$ denote the eigenvalues of the plotted block $P\equiv P_+$. The complete third-quantized structure matrix contains two $N_s$-dimensional blocks,
\begin{equation}
	P_+=h+\frac{i\gamma}{2}\Pi_s,
	\qquad
	P_-=-h+\frac{i\gamma}{2}\Pi_s.
\end{equation}
For the SSH chain, the chiral operator $\Sigma$ obeys $\Sigma P_+\Sigma=P_-$, so the two blocks are isospectral. Consequently, each eigenvalue $E_j$ of $P$ appears once in each block, and the complete set of $2N_s$ elementary Liouvillian rapidities is
\begin{equation}
\beta_j^{(+)}=\beta_j^{(-)}\equiv iE_j,
\qquad j=1,\ldots,N_s.
\label{eq:rapidities_relation}
\end{equation}
A rapidity is the complex time-evolution exponent of one elementary normal master mode: a mode with rapidity $\beta_j^{(\sigma)}$ contributes a factor $e^{\beta_j^{(\sigma)}t}$. Hence $-\Re\beta_j^{(\sigma)}=\Im E_j$ is its decay rate, while $\Im\beta_j^{(\sigma)}=\Re E_j$ is its oscillation frequency. Fermionic subset sums of the complete $2N_s$-rapidity multiset generate all $2^{2N_s}=4^{N_s}$ Liouvillian eigenvalues. The ``rapidity edge'' is therefore the least-damped elementary edge, and ``rapidity reorganization'' denotes its spectral rearrangement as $\gamma$ is varied. For the complete operator space, the smallest positive elementary decay rate fixes the standard Liouvillian gap,
\begin{equation}
\Delta_{\mathcal L}=\min_{j:\,\Im E_j>0}\Im E_j.
\label{eq:gap_from_reduction}
\end{equation}
Parity-even density matrices and observables sample even subset sums of the rapidities; their sector-resolved decay edge is discussed explicitly in Appendix~\ref{app:third_quantization}. The many-body relaxation problem is thus reduced exactly to the spectrum of the rank-one impurity matrix in Eq.~\eqref{eq:P_imp_def}.

\subsection{Statistics dependence and experimental observables}

\label{subsec:statistics_experiment}

It is useful to separate two logically distinct statements about the reduction. First, Eqs.~\eqref{eq:jump_bulk}--\eqref{eq:gap_from_reduction} define and solve a genuine fermionic many-body Lindblad problem: third quantization reconstructs its full $4^{N_s}$-dimensional operator-space spectrum, including the even and odd operator-parity sectors and the fermionic occupation constraints $n_{j,\sigma}=0,1$ on normal master modes. This part is not a single-particle approximation. Second, because the Hamiltonian is quadratic and the loss operator is linear, all distinct elementary decay exponents are fixed by the one-body drift matrix $M=-ih-\Gamma/2$, or equivalently can be obtained from the sign-conjugated spectral block $P=h+i\Gamma/2$ used below. Consequently, the weak-loss local-weight rule, exceptional-point coalescence, and strong-loss Zeno elimination are Gaussian linear mechanisms and do not rely on Pauli blocking or on fermionic exchange statistics.

The same matrix therefore governs several experimentally different realizations, but the level of correspondence must be stated carefully. In a noninteracting fermionic cold-atom chain, the full fermionic Lindblad formulation applies, and site-resolved measurements can access $n_j(t)$, the surviving particle number, and window observables such as $W_L(t)$~\cite{Cooper2019,meier2016}. In a linear bosonic resonator or photonic array with local loss, the coherent amplitudes obey $\dot{\bm\alpha}=M\bm\alpha$, whereas the one-body correlation matrix obeys $\dot C=MC+CM^\dagger$. Both are therefore controlled by the same non-Hermitian drift matrix, so the rapidity spectrum can be inferred from resonance positions and linewidths, transmission spectroscopy, or time-domain measurements of local intensities~\cite{Blais2021,Ozawa2019}. A classical coupled-mode network realizes the same non-Hermitian eigenvalue problem for field amplitudes and can reproduce the local-weight selection, exceptional points, and Zeno cut-chain crossover. Such bosonic and classical platforms do not, by themselves, reproduce the finite fermionic operator-parity sectors or the $0/1$ subset-sum organization of the complete many-body Liouvillian. Thus the rapidity mechanisms are statistics independent at the linear level, whereas the reconstruction and sector organization of the full Liouvillian spectrum are specifically fermionic.

Experimentally, reconstruction of the entire many-body Liouvillian is not required to test the main spectral predictions. The weak-loss slopes follow from mode linewidths or local density decay, the exceptional point is identified by eigenfrequency coalescence and vanishing phase rigidity, and the Zeno regime is revealed by suppression of occupation on the lossy site, recovery of the two cut-chain mode families, and the asymptotic $1/\gamma$ relaxation scale. The exact one-body observables used in Fig.~\ref{fig:intermediate_gamma} were chosen for precisely this reason: they are directly measurable in fermionic quantum simulators and have immediate amplitude- or intensity-level counterparts in linear bosonic and classical-wave platforms.

\section{Exact rapidity equation for a single bulk imaginary impurity}

\label{sec:exact_rapidity}

We next derive a closed secular equation for the eigenvalues $E$ of the $2N\times 2N$ non-Hermitian matrix $P$. The corresponding rapidities are parameterized as $\beta=iE$ according to Eq.~\eqref{eq:rapidities_relation}. The matrix $P$ is defined
as 
\begin{equation}
P=P_{0}+\frac{i\gamma}{2}\,\Pi _{s},\qquad \Pi _{s}\equiv \bm e_{s}\bm %
e_{s}^{T},\qquad \gamma \geq 0,  \label{eq:P_rank1_main}
\end{equation}%
where $P_{0}$ is the clean SSH tridiagonal hopping matrix with vanishing
diagonal entries and $\bm e_{s}$ is the standard basis vector on the
impurity site $s$. For a particle loss placed on the A site of cell $m$, the
impurity index in the ordered basis $(1A,1B,2A,2B,\dots ,NA,NB)$ is $s=2m-1$.

\subsection{Rank-one reduction to a scalar secular equation}

\label{subsec:rank_one_secular}

The eigenvalues $E$ satisfy the characteristic condition $\det(E\boldsymbol{I%
}_{2N}-P)=0$. Substituting Eq.~\eqref{eq:P_rank1_main} yields 
\begin{equation}
E\boldsymbol{I}_{2N}-P=\big(E\boldsymbol{I}_{2N}-P_0\big)-\frac{i\gamma}{2}\,\bm e_s
\bm e_s^{T}.  \label{eq:A_def}
\end{equation}
This expression represents a rank-one update of the clean system matrix.
Provided the clean matrix $E\boldsymbol{I}_{2N}-P_0$ is invertible, which
holds for $E$ outside the spectrum of $P_0$, the matrix determinant lemma
gives 
\begin{equation}
\begin{split}
\det\!\big(E\boldsymbol{I}_{2N}-P\big)
&= \det\!\big(E\boldsymbol{I}_{2N}-P_0\big) \\
&\quad \times \left[1-\frac{i\gamma}{2}\,\bm e_s^{T}(E\boldsymbol{I}_{2N}-P_0)^{-1}\bm e_s\right].
\end{split}
\label{eq:det_factorized}
\end{equation}
The local resolvent of the clean chain is defined as 
\begin{equation}
G_0(s,s;E)\equiv \bm e_s^{T}\big(E\boldsymbol{I}_{2N}-P_0\big)^{-1}\bm e_s.
\label{eq:G0_local_def}
\end{equation}
The eigenvalues of $P$ are consequently determined by the scalar equation 
\begin{equation}
1-\frac{i\gamma}{2}\,G_0(s,s;E)=0.  \label{eq:scalar_secular}
\end{equation}
For roots with $E\notin\operatorname{spec}(P_0)$, Eq.~\eqref{eq:scalar_secular}
determines the impurity-coupled eigenvalues of the finite rapidity matrix $P$.
If a clean eigenmode satisfies
$P_0\bm\phi_\alpha=E_\alpha^{(0)}\bm\phi_\alpha$ and
$\phi_\alpha(s)=0$, then $\Pi_s\bm\phi_\alpha=0$ and it remains an exact dark
eigenmode of $P$ at the unchanged eigenvalue $E_\alpha^{(0)}$. Such a root is
not represented as an ordinary zero of Eq.~\eqref{eq:scalar_secular}, because
the resolvent form assumes that
$E\boldsymbol{I}_{2N}-P_0$ is invertible. The unconditional characteristic
polynomial, which contains both impurity-coupled roots and any exact dark roots,
is Eq.~\eqref{eq:secular_D} below. The full many-body Liouvillian spectrum is
then reconstructed from fermionic subset sums of the two rapidity blocks, as
derived in Appendix~\ref{app:third_quantization}.

\subsection{Resolvent as a ratio of tridiagonal determinants}

\label{subsec:resolvent_ratio}

Cramer's rule expresses the diagonal resolvent element as a ratio of
determinants, 
\begin{equation}
G_{0}(s,s;E)=\frac{\det \!\big(E\boldsymbol{I}_{2N-1}-P_{0}^{(s)}\big)}{\det %
\big(E\boldsymbol{I}_{2N}-P_{0}\big)},  \label{eq:G0_ratio}
\end{equation}%
where $P_{0}^{(s)}$ is the minor obtained by removing the $s$-th row and
column from $P_{0}$. The tridiagonal structure of $P_{0}$ implies that the
removal of site $s$ decouples the chain into two independent segments of
lengths $s-1$ and $2N-s$. The minor determinant therefore factorizes into
the product of the determinants of these sub-segments.

Segment determinants are defined as $D_{\ell}(E)\equiv \det(E\boldsymbol{I}%
_{\ell}-P_{0,\ell})$, where $P_{0,\ell}$ is the clean SSH matrix restricted
to the first $\ell$ sites. The factorization of the minor leads to 
\begin{equation}
\det\!\big(E\boldsymbol{I}_{2N-1}-P_0^{(s)}\big)=D_{s-1}(E)\,D_{2N-s}(E),
\label{eq:minor_factor}
\end{equation}
which simplifies the local resolvent to the form 
\begin{equation}
G_0(s,s;E)=\frac{D_{s-1}(E)\,D_{2N-s}(E)}{D_{2N}(E)}.  \label{eq:G0_D_ratio}
\end{equation}
Inserting this result into Eq.~\eqref{eq:scalar_secular} yields the exact
secular equation 
\begin{equation}
D_{2N}(E)-\frac{i\gamma}{2}\,D_{s-1}(E)\,D_{2N-s}(E)=0.  \label{eq:secular_D}
\end{equation}

\subsection{Closed forms and Chebyshev representation}

\label{subsec:D_closed_forms}

The tridiagonal structure of $P_0$ with alternating couplings $J_{2n-1}=t_1$
and $J_{2n}=t_2$ dictates a specific recursion for the determinants. The
determinant $D_{\ell}(E)$ obeys 
\begin{equation}
D_{\ell}(E)=E\,D_{\ell-1}(E)-J_{\ell-1}^{2}\,D_{\ell-2}(E),  \label{eq:D_rec}
\end{equation}
with initial conditions $D_0=1$ and $D_1=E$. Separating the even-length and
odd-length determinants, $F_n(E)\equiv D_{2n}(E)$ and $O_n(E)\equiv
D_{2n+1}(E)$, reduces the problem to a constant-coefficient recursion 
\begin{equation}
Y_n=(E^2-t_1^2-t_2^2)\,Y_{n-1}-(t_1 t_2)^2\,Y_{n-2}.  \label{eq:Y_rec}
\end{equation}
The dimensionless spectral variable is defined as $x =
(E^2-t_1^2-t_2^2)/(2t_1 t_2)$. This identification allows the determinants
to be expressed in terms of Chebyshev polynomials of the second kind $U_n(x)$
as 
\begin{align}
D_{2n}(E)&=(t_1 t_2)^n\left[U_n(x)+\frac{t_2}{t_1}U_{n-1}(x)\right],
\label{eq:D2n_closed} \\
D_{2n+1}(E)&=E\,(t_1 t_2)^n\,U_n(x).  \label{eq:D2n1_closed}
\end{align}

For a loss localized on the site $s=2m-1$, the segment lengths are $%
s-1=2(m-1)$ and $2N-s=2(N-m)+1$. The secular equation in Eq.~%
\eqref{eq:secular_D} becomes 
\begin{equation}
F_{N}(E)-\frac{i\gamma}{2}\,F_{m-1}(E)\,O_{N-m}(E)=0.  \label{eq:secular_FO}
\end{equation}%
Substituting the closed forms and rescaling by $(t_{1}t_{2})^{N-1}$ produces
the final exact rapidity equation 
\begin{multline}
t_{1}t_{2}\!\left[ U_{N}(x)+\frac{t_{2}}{t_{1}}U_{N-1}(x)\right] \\
- \frac{i\gamma}{2}\,E\,U_{N-m}(x)\!\left[ U_{m-1}(x)+\frac{t_{2}}{t_{1}}U_{m-2}(x)\right] = 0.
\label{eq:secular_final}
\end{multline}
Eq.~\eqref{eq:secular_final} provides an exact analytical starting point for
the complex spectrum across all dissipation regimes. It allows the
Liouvillian gap and the rapidity reorganization to be analyzed directly. 
%========================================================
%========================================================

%========================================================

\section{Impurity-induced bound rapidities and eigenvectors}

\label{sec:bound_rapidities}

Before turning to the asymptotic loss regimes, it is useful to extract from Eq.~\eqref{eq:secular_final} the types of rapidities it supports and the corresponding eigenvector structure. In the topological regime $|t_{1}|<|t_{2}|$, the clean SSH matrix $P_{0}$ supports two exponentially localized edge modes with energies exponentially close to zero. After a bulk loss is introduced at site $s=2m-1$, these modes remain edge localized, and their decay rates are set by the edge-mode weight on the lossy site, $|\psi_{\mathrm{edge}}(s)|^{2}$. This overlap scale controls the weak-$\gamma$ expansion below.

The rank-one imaginary impurity does not add an extra eigenvalue to the fixed-dimensional spectrum. Instead, as the loss is increased, one existing spectral branch progressively acquires strong weight on the lossy site. Its decay is then governed directly by the amplitude-loss scale $\gamma/2$, and in the limit $\gamma\gg |t_{1,2}|$ it evolves into the ultrafast root $E\simeq i\gamma/2$ and becomes localized almost entirely on site $s$. This branch does not control the slowest relaxation at weak loss, but it becomes central in the strong-loss regime. Once it separates from the slow sector, the lossy site acts effectively as a cut, and the remaining long-lived spectrum is reorganized on the two disconnected segments. This is the origin of the geometry-dependent scaling of the Liouvillian gap derived below.

\subsection{Bulk-band parameterization and analytic continuation}

\label{subsec:bulk_param}

The Chebyshev polynomials of the second kind satisfy 
\begin{equation}
U_{n}(\cos k)=\frac{\sin[(n+1)k]}{\sin k},\qquad k\in\mathbb{C}.
\label{eq:Un_sin}
\end{equation}
The spectral variable entering \eqref{eq:secular_final} is 
\begin{equation}
x\equiv \frac{E^2-t_1^2-t_2^2}{2t_1 t_2}.  \label{eq:x_def_bound}
\end{equation}
We parameterize $x$ by a generally complex quasi-momentum $k$ according to 
\begin{equation}
x=\cos k \quad \Longleftrightarrow \quad
E^{2}=t_{1}^{2}+t_{2}^{2}+2t_{1}t_{2}\cos k .  \label{eq:dispersion_k}
\end{equation}
For real $k\in(0,\pi)$, Eq.~\eqref{eq:dispersion_k} reproduces the clean SSH
bulk dispersion, with the two branches $\pm E$. Localized solutions are
obtained for $|x|>1$, which corresponds to complex $k$.

A convenient continuation is $x=\cosh\kappa$ with $\kappa>0$, corresponding
to $k=i\kappa$ or $k=\pi+i\kappa$. In this case, 
\begin{equation}
U_{n}(\cosh \kappa)=\frac{\sinh[(n+1)\kappa]}{\sinh\kappa}.
\label{eq:Un_sinh}
\end{equation}
The resulting exponential dependence on $n$ is the origin of spatial
localization for both impurity-localized and edge-localized states.

\subsection{Eigenvector construction and localization}

\label{subsec:eigenvector_localization}

Once an eigenvalue $E$ of $P$, equivalently a rapidity $\beta=iE$, is
obtained from the exact secular equation \eqref{eq:scalar_secular} or %
\eqref{eq:secular_final}, the corresponding right eigenvector can be
constructed directly from the clean resolvent.

Starting from 
\begin{equation}
(E\boldsymbol{I}_{2N}-P)\,\bm\psi=\bm 0,  \label{eq:eig_eq_vec}
\end{equation}
and using $P=P_0+i(\gamma/2)\Pi_s$ with $\Pi_s=\bm e_s\bm e_s^{T}$, one finds 
\begin{equation}
(E\boldsymbol{I}_{2N}-P_0)\,\bm\psi = \frac{i\gamma}{2}\,(\bm e_s^{T}\bm\psi)\,\bm e_s
= \frac{i\gamma}{2}\,\psi_s\,\bm e_s,  \label{eq:eig_eq_rearranged_vec}
\end{equation}
where $\psi_s\equiv \psi(s)$ denotes the amplitude on the lossy site.
Provided that $E\notin \mathrm{spec}(P_0)$, the operator $(E\boldsymbol{I}%
_{2N}-P_0)$ is invertible, and therefore 
\begin{equation}
\bm\psi = \frac{i\gamma}{2}\,\psi_s\,(E\boldsymbol{I}_{2N}-P_0)^{-1}\bm e_s.
\label{eq:psi_resolvent_vec}
\end{equation}
Since $E$ already satisfies the secular equation, the scalar factor $\psi_s$
is fixed up to an overall normalization. The right eigenvector is thus 
\begin{equation}
\bm\psi \propto (E\boldsymbol{I}_{2N}-P_0)^{-1}\bm e_s.
\label{eq:eigvector_final}
\end{equation}

Equation \eqref{eq:eigvector_final} directly relates the spatial structure
of the eigenvector to the analytic structure of the resolvent. Writing $%
x=\cos k$ through \eqref{eq:dispersion_k}, one finds that real $k$ gives an
extended bulk state, whereas complex $k$ yields an exponentially decaying
resolvent kernel and hence a localized state. For $k=i\kappa$ with $\kappa>0$%
, the envelope takes the form 
\begin{equation}
|\psi_j|\sim e^{-|j-s|/\xi}, \qquad \xi^{-1}\equiv \kappa.  \label{eq:xi_def}
\end{equation}

For the impurity-localized mode, one typically has $|\psi_s|=O(1)$, so its
decay rate is set by $\gamma/2$ at the amplitude level and it mainly governs the short-time density
depletion around the lossy site. In the Zeno regime $\gamma\gg |t_{1,2}|$,
this mode approaches the ultrafast root $E\simeq i\gamma/2$ and separates from
the slow sector. The remaining long-lived modes are then governed by the two
segments on the left and right of the lossy site. By contrast, in the
topological regime $|t_{1}|<|t_{2}|$, the two edge modes carry only
exponentially small weight on the bulk site $s$. Their decay rates are
therefore parametrically smaller and control the long-time Liouvillian
dynamics at weak loss through the overlap $|\psi_{\mathrm{edge}}(s)|^{2}$.

\section{Small-loss regime: local spectral weight and the weak Liouvillian
gap}

\label{sec:weak_gamma}

We now turn to the weak-loss regime $\gamma\to 0$. In this limit, the exact impurity problem reduces to a controlled perturbation around the clean SSH spectrum, so the Liouvillian gap can be traced directly to the local spectral weight of the clean modes at the lossy site.

Let $P_0$ be the clean Hermitian single-particle matrix with normalized
eigenpairs 
\begin{equation}
P_{0}\,\bm\phi_\alpha = E_{\alpha }^{(0)}\,\bm\phi_\alpha, \qquad \bm%
\phi_\alpha^\dagger \bm\phi_\beta=\delta_{\alpha\beta}.
\label{eq:clean_spectrum_weak}
\end{equation}
A single bulk loss at lattice site $s$ enters as 
\begin{equation}
P=P_{0}+\frac{i\gamma}{2}\,\Pi_s, \qquad \Pi_s\equiv \bm e_s \bm e_s^{T}, \qquad
\gamma>0,  \label{eq:P_def_weak}
\end{equation}
where $\bm e_s$ is the standard basis vector at the lossy site. For a loss
on $(m,A)$, one has $s=2m-1$. Within our convention, the Liouvillian
rapidities are $\beta_\alpha=iE_\alpha$, so $-\Re\beta_\alpha=\Im E_\alpha$. The weak-loss Liouvillian gap is therefore set by the smallest
nonzero $\Im E_\alpha$.

From Sec.~\ref{sec:exact_rapidity}, the eigenvalues of $P=P_{0}+i(\gamma/2)\Pi_s$ satisfy the exact scalar equation 
\begin{equation}
1-\frac{i\gamma}{2}\,G_{0}(s,s;E)=0.  \label{eq:scalar_secular_weak}
\end{equation}%
We now solve Eq.~\eqref{eq:scalar_secular_weak} perturbatively near a clean
eigenvalue. Using Eq.~\eqref{eq:clean_spectrum_weak}, the clean resolvent
has the spectral representation 
\begin{equation}
G_{0}(E)=\sum_{\beta }\frac{\bm\phi _{\beta }\bm\phi _{\beta }^{\dagger }}{%
E-E_{\beta }^{(0)}},\qquad G_{0}(s,s;E)=\sum_{\beta }\frac{|\phi _{\beta
}(s)|^{2}}{E-E_{\beta }^{(0)}},  \label{eq:G0_spectral_weak}
\end{equation}%
where $\phi _{\beta }(s)\equiv \bm e_{s}^{T}\bm\phi _{\beta }$ is the clean
wave-function amplitude at the lossy site. Fix a mode $\alpha $ with $\phi
_{\alpha }(s)\neq 0$ and write 
\begin{equation}
E_{\alpha }=E_{\alpha }^{(0)}+\delta E_{\alpha },\qquad |\delta E_{\alpha
}|\ll 1.  \label{eq:E_shift_ansatz_weak}
\end{equation}%
Near $E_{\alpha }^{(0)}$, we separate the singular and regular parts as 
\begin{equation}
G_{0}(s,s;E)=\frac{|\phi _{\alpha }(s)|^{2}}{E-E_{\alpha }^{(0)}}+R_{\alpha
}(s;E),  \label{eq:pole_plus_regular}
\end{equation}%
with 
\begin{equation}
R_{\alpha }(s;E)\equiv \sum_{\beta \neq \alpha }\frac{|\phi _{\beta }(s)|^{2}%
}{E-E_{\beta }^{(0)}}.  \label{eq:Ralpha_def}
\end{equation}%
Since $R_{\alpha }(s;E)$ is analytic at $E=E_{\alpha }^{(0)}$, one has 
\begin{equation}
R_{\alpha }\!\bigl(s;E_{\alpha }^{(0)}+\delta E_{\alpha }\bigr)=R_{\alpha }\!%
\bigl(s;E_{\alpha }^{(0)}\bigr)+O(\delta E_{\alpha }).
\label{eq:Ralpha_Taylor_weak}
\end{equation}%
Substituting Eqs.~\eqref{eq:E_shift_ansatz_weak}--%
\eqref{eq:Ralpha_Taylor_weak} into Eq.~\eqref{eq:scalar_secular_weak} gives 
\begin{equation}
1-\frac{i\gamma}{2} \left[ \frac{|\phi _{\alpha }(s)|^{2}}{\delta E_{\alpha }}%
+R_{\alpha }\!\bigl(s;E_{\alpha }^{(0)}\bigr)+O(\delta E_{\alpha })\right]
=0.  \label{eq:scalar_expanded}
\end{equation}%
Multiplying by $\delta E_{\alpha }$, we obtain 
\begin{equation}
\delta E_{\alpha }-\frac{i\gamma}{2}|\phi _{\alpha }(s)|^{2}-\frac{i\gamma}{2}\,\delta
E_{\alpha }\,R_{\alpha }\!\bigl(s;E_{\alpha }^{(0)}\bigr)+O(\gamma \,\delta
E_{\alpha }^{2})=0.  \label{eq:deltaE_intermediate_weak}
\end{equation}%
Self-consistency requires $\delta E_{\alpha }=O(\gamma )$, so the last term
is $O(\gamma ^{3})$ and does not enter at leading order. Therefore  
\begin{equation}
E_{\alpha }=E_{\alpha }^{(0)}+\frac{i\gamma}{2}\,|\phi _{\alpha }(s)|^{2}+O(\gamma
^{2}).  \label{eq:E_shift_final_weak}
\end{equation}%
Each clean mode with nonzero weight on the lossy site thus acquires an
imaginary part proportional to its local spectral weight. Using $\beta_{\alpha }=iE_{\alpha }$, one finds 
\begin{equation}
-\Re \beta _{\alpha }=\frac{\gamma}{2}\,|\phi _{\alpha }(s)|^{2}+O(\gamma ^{2}).
\label{eq:rapidity_shift_final_weak}
\end{equation}%
If a clean eigenmode has a node at the dissipative site, $\phi _{\alpha
}(s)=0$, then $\Pi _{s}\bm\phi _{\alpha }=\bm0$, so $\bm\phi _{\alpha }$
remains an exact eigenvector of $P$ with unchanged eigenvalue $E_{\alpha
}=E_{\alpha }^{(0)}$. Such modes are decoupled from this loss channel and do
not contribute to the leading weak-$\gamma $ Liouvillian gap.

\subsection{Weak-\texorpdfstring{$\gamma$}{gamma} Liouvillian gap and scaling laws}

\label{subsec:weak_gap_scaling}

The weak-loss Liouvillian gap is the smallest nonzero decay rate, 
\begin{equation}
\Delta _{\mathcal{L}}\equiv \frac{\gamma}{2}\,\min_{\alpha :\phi _{\alpha }(s)\neq
0}|\phi _{\alpha }(s)|^{2}+O(\gamma ^{2}).  \label{eq:gap_weak_general}
\end{equation}%
The problem is therefore reduced to determining the smallest clean-site
weight among the slow modes.

We first consider the generic case in which the slow sector is formed by
extended standing waves. For open boundaries, the quantized momenta satisfy 
\begin{equation}
k_{m}\sim \frac{m\pi }{N+1},\qquad m=1,2,\dots ,  \label{eq:k_quant_gap}
\end{equation}%
so the smallest momentum scales as $k_{\min }\sim \pi /(N+1)$. Near a
boundary, the standing-wave envelope behaves as $\psi _{k}(n)\propto \sin
(nk)$, and therefore $\psi _{k}(1)\propto \sin k\sim k$ for $k\rightarrow 0$%
. Since an extended state normalized over $O(N)$ unit cells has typical
magnitude $|\psi _{k}|\sim N^{-1/2}$, the weight of the slowest
boundary-sensitive mode satisfies 
\begin{equation}
|\psi _{k_{\min }}(s)|^{2}\sim \left( \frac{\sin k_{\min }}{\sqrt{N}}\right)
^{2}\sim \frac{k_{\min }^{2}}{N}\sim \frac{1}{N^{3}}.
\label{eq:boundary_weight_Nm3_gap}
\end{equation}%
Substituting this estimate into Eq.~\eqref{eq:gap_weak_general} yields $%
\Delta _{\mathcal{L}}(N)\asymp (\gamma/2) N^{-3}$ with $\gamma \ll
|t_{1}|,|t_{2}|$, for a generic boundary loss. The same envelope typically
persists for a defect at a generic bulk site. Indeed, a standing-wave
estimate gives $|\phi _{\alpha }(s)|^{2}\sim \frac{1}{N}\sin ^{2}(k_{\alpha
}s)$. Since the open-boundary momentum spacing scales as $\Delta k\sim 1/N$,
one can typically find a mode whose node lies within $O(1/N)$ of site $s$.
This gives $\sin ^{2}(k_{\alpha }s)\sim N^{-2}$ and hence
\begin{equation}
\Delta _{\mathcal{L}}^{\mathrm{(bulk)}}\sim \frac{\gamma}{2N^{3}},
\label{eq:gap_bulk_weak}
\end{equation}
up to an $O(1)$ prefactor determined by the commensurability between $s$ and the open-boundary momentum grid. At special commensurate positions, exact nodes may occur for some bulk modes, in which case the gap is set by the next slowest mode that remains coupled to the defect.

We next turn to the topological regime $|t_{1}|<|t_{2}|$, where $P_{0}$
supports exponentially localized edge modes. Their overlap with a bulk loss
can be exponentially small, and the weak-loss gap then becomes edge
controlled. The localization parameter is defined as $r\equiv \left\vert 
\frac{t_{1}}{t_{2}}\right\vert <1$. In our convention, the left edge mode in
the semi-infinite limit lives predominantly on the $A$ sublattice and
satisfies the zero-energy recursion $t_{1}a_{n}+t_{2}a_{n+1}=0$ so that $%
|a_{n}|\propto r^{\,n-1}$. Normalizing $\sum_{n=1}^{N}|a_{n}|^{2}=1$ gives 
\begin{equation}
|a_{m}|^{2}=\frac{(1-r^{2})\,r^{2(m-1)}}{1-r^{2N}}\simeq
(1-r^{2})\,r^{2(m-1)}\qquad (N\gg 1).  \label{eq:edge_weight_at_m}
\end{equation}%
For a loss on the $A$ site of cell $m$, namely $s=2m-1$, the overlap with
the left edge mode therefore satisfies $|\phi _{\mathrm{L}}(m,A)|^{2}\simeq
(1-r^{2})\,r^{2(m-1)}$, and Eq.~\eqref{eq:gap_weak_general} gives 
\begin{equation}
\Delta _{\mathcal{L}}^{\mathrm{(edge)}}\sim \frac{\gamma}{2} \,|\phi _{\mathrm{L}%
}(m,A)|^{2}\sim \frac{\gamma}{2}(1-r^{2})\,r^{2(m-1)}.  \label{eq:gap_edge_weak}
\end{equation}%
The weak-loss gap therefore depends exponentially on the distance between
the lossy site and the boundary.

\begin{figure*}[t]
\centering
\includegraphics[width=\textwidth]{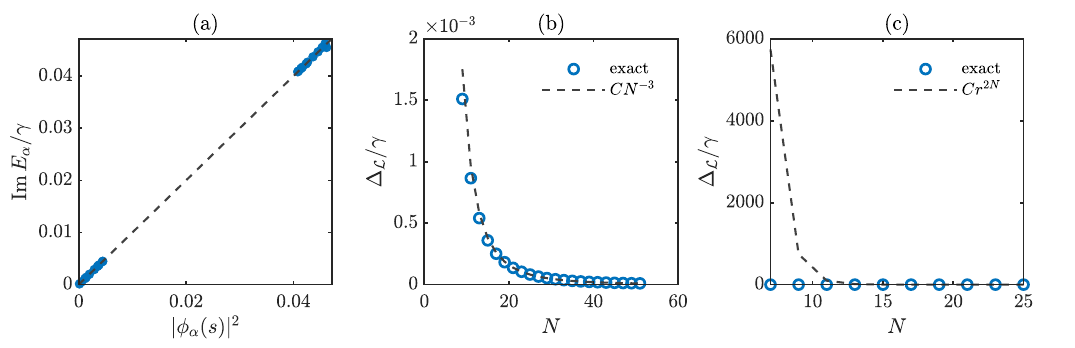}%
\caption{Weak-loss regime. (a) For $N=21$, $t_1=1.0$, $t_2=0.6$, $m=11$, and $\gamma=0.05$, the exact rapidity shifts satisfy $\Im E_\alpha/(\gamma/2)\approx |\phi_\alpha(s)|^2$, in agreement with Eq.~\eqref{eq:E_shift_final_weak}. Here $|\phi_\alpha(s)|^2$ is the clean local spectral weight at the lossy site. (b) For a trivial chain with $t_1=1.0$, $t_2=0.6$, $\gamma=0.05$, and $m=(N+1)/2$, the rescaled gap $2\Delta_{\mathcal L}/\gamma$ follows the generic law $N^{-3}$ as $N$ is varied. Only odd sizes $N=9,11,\ldots,51$ are included, so that the dissipative site is always the central unit cell and no two points correspond to the same system size. (c) For a topological chain with $t_1=0.6$, $t_2=1.0$, and $\gamma=0.05$, the rescaled gap $2\Delta_{\mathcal L}/\gamma$ is plotted against $N$ for a boundary loss $L=\sqrt{\gamma}\,c_{1,B}$. The data follow the exponentially small boundary law $r^{2N}$ with $r=|t_1/t_2|$, in agreement with Eq.~\eqref{eq:gap_weak_exp_gap}. In panels (b) and (c), the dashed lines are guides to the eye of the form $C N^{-3}$ and $C' r^{2N}$, respectively, with $C$ and $C'$ independent of $N$.}
\label{fig:small_gamma}
\end{figure*}

For a finite chain, the left and right edge states hybridize through their
exponentially small overlap, with amplitude of order $r^{N}$. For a boundary loss $L=\sqrt{\gamma}\,c_{1,B}$, the lossy site couples to the exponentially small tail of the opposite edge mode. Consequently, $|\psi_{\mathrm{edge}}(1B)|^{2}\sim r^{2N}$, which implies
\begin{equation}
\Delta _{\mathcal{L}}(N)\asymp \frac{\gamma}{2} r^{2N}
\qquad
(\gamma \ll |t_{1}|,|t_{2}|,\quad |t_{1}|<|t_{2}|).
\label{eq:gap_weak_exp_gap}
\end{equation}
These edge-controlled estimates refer to different loss geometries and should not be combined into one universal minimum. For a bulk loss on the $A$ site of cell $m$, the relevant weak-loss competition is
\begin{equation}
\Delta_{\mathcal L}^{\rm bulk\ loss}
\sim \frac{\gamma}{2}\min\left\{
(1-r^2)r^{2(m-1)},\,\frac{c(s)}{N^3}
\right\},
\label{eq:gap_weak_combined}
\end{equation}
where $c(s)=O(1)$ describes the position and commensurability dependence of the extended-mode envelope. By contrast, for the boundary-$B$ loss used in Fig.~\ref{fig:small_gamma}(c), the opposite-edge tail gives the separate law in Eq.~\eqref{eq:gap_weak_exp_gap}, $\Delta_{\mathcal L}\asymp(\gamma/2)r^{2N}$.

Because $P(\gamma)$ has fixed dimension and depends analytically on $\gamma$, every finite eigenvalue approaches the clean spectrum as $\gamma\to0$. At finite loss, however, one branch can progressively acquire strong weight on the lossy site and evolve into the impurity-localized rapidity that becomes ultrafast at large $\gamma$. Its behavior over the full crossover is not captured by a local single-pole expansion about one fixed clean eigenvalue. At weak loss this branch has a decay scale of order $\gamma/2$ and therefore does not determine $\Delta_{\mathcal L}$, although it can affect short-time local dynamics and becomes important in the strong-loss reorganization discussed later.

These estimates identify the weak-loss hierarchy of decay channels directly from the clean eigenstates. Figure~\ref{fig:small_gamma} summarizes the weak-loss regime with three complementary diagnostics. With the physical Lindblad rate $\gamma$, the perturbative relation is $\Im E_\alpha/(\gamma/2) \approx |\phi_\alpha(s)|^2$, and the natural weak-loss rescaling is $2\Delta_{\mathcal L}/\gamma$. Panel~(a) uses a trivial chain with $N=21$, $t_1=1.0$, $t_2=0.6$, $m=11$, and $\gamma=0.05$ to test this relation. Panel~(b) then varies $N$ at fixed $\gamma=0.05$ and $m=(N+1)/2$ in the same trivial regime and shows the generic algebraic law $2\Delta_{\mathcal{L}}/\gamma\propto N^{-3}$. Only odd sizes $N=9,11,\ldots,51$ are used, so that $m=(N+1)/2$ is an integer and the loss remains at the central unit cell. Panel~(c) turns to the topological regime with $t_1=0.6$, $t_2=1.0$, and $\gamma=0.05$, but now uses a boundary loss $L=\sqrt{\gamma}\,c_{1,B}$ and varies $N$. In this geometry the gap is controlled by the exponentially weak hybridization of the opposite edge mode, and the numerical data follow the boundary law $2\Delta_{\mathcal{L}}/\gamma\propto r^{2N}$. The three panels therefore isolate the perturbative local-weight rule, the generic bulk $N^{-3}$ law, and the exponentially small boundary-controlled gap in the topological phase.

\section{Intermediate loss regime: rapidity reorganization,
\texorpdfstring{exact-EP dynamics}{exact-EP dynamics}, and slow relaxation}

\label{sec:intermediate_gamma}

\begin{figure*}[t]
\centering
\includegraphics[width=\textwidth]{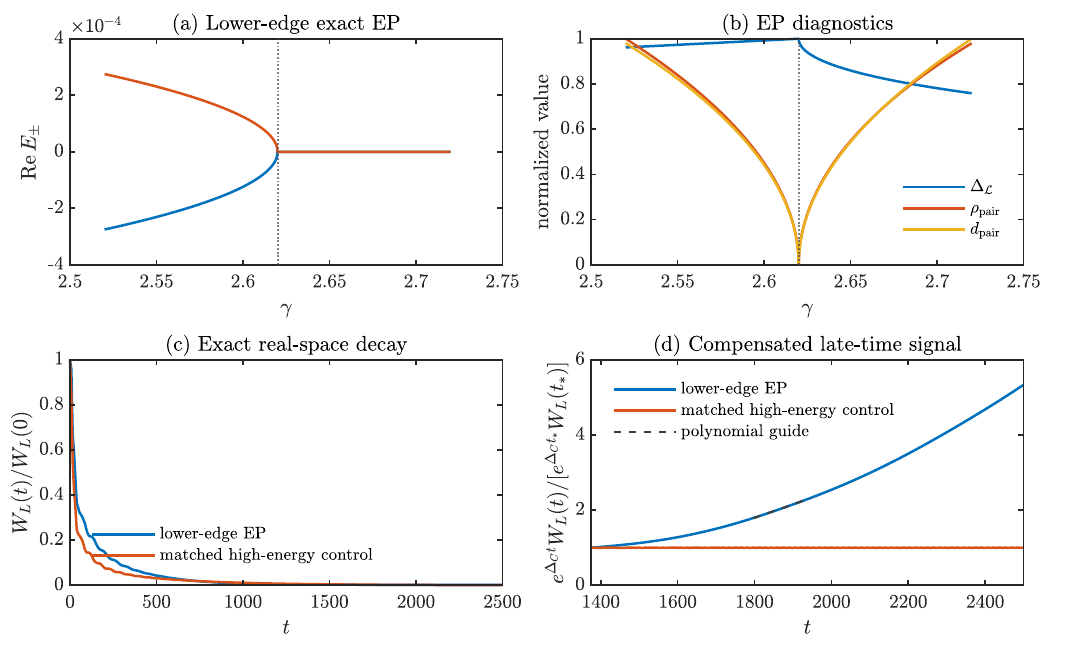}%
\caption{Intermediate-loss regime and its real-space signature. Panels (a) and (b) diagnose an exact exceptional point reached on the real-$\gamma$ axis for a chain with $N=15$, $t_2=1.0$, and loss on $(m,A)=(8,A)$ at fixed $t_1=0.655$. Panel (a) tracks the lower-edge rapidity pair and shows its coalescence at $\gamma_{\mathrm{EP}}=2.6200$. The centered-loss symmetry produces simultaneous coalescences of the other chiral-partner pairs; the displayed pair is singled out because it fixes the lower rapidity edge. Panel (b) compares the normalized quantities $\Delta_{\mathcal L}(\gamma)$, $\rho_{\rm pair}(\gamma)$, and $d_{\rm pair}(\gamma)$ defined in Eqs.~\eqref{eq:rho_pair_def} and \eqref{eq:d_pair_def}. Their simultaneous vanishing at the gap cusp establishes exact defectiveness of the tracked lower-edge pair. Panels (c) and (d) show the exact one-body dynamics for an initial excitation on site $(3,A)$. The lower-edge EP $(t_1,\gamma)=(0.655,2.6200)$ and the matched high-energy control $(0.8465,2.2421)$ both have $\Delta_C\simeq2.005648\times10^{-3}$ and $\Delta_{\mathcal L}\simeq1.002824\times10^{-3}$. Panel (c) plots $W_L(t)/W_L(0)$, and panel (d) plots the compensated and anchored signal $\widetilde W_L(t)$ defined in Eq.~\eqref{eq:WL_compensated_main}. The short dashed segment is a local linear fit over a restricted late-time subwindow and serves only as a visual polynomial guide. Only the lower-edge EP develops the pronounced polynomially enhanced late-time prefactor, while the matched control remains nearly flat.}
\label{fig:intermediate_gamma}
\end{figure*}

The weak-loss analysis shows how decay channels first open from the clean spectrum. We now turn to the intermediate regime, where perturbation theory is no longer sufficient and nearby rapidity branches can reorganize over a narrow loss window. In a finite chain, this reorganization is often strongest near the lower edge of the rapidity spectrum, where it can directly reshape the Liouvillian gap. We reserve the term exceptional point for an actual double root reached on the real-$\gamma$ axis; a narrow nearest approach with suppressed phase rigidity but finite pair distance is referred to as near-EP behavior. For the centered $A$-sublattice loss used in Fig.~\ref{fig:intermediate_gamma}, the chiral-partner weights are exactly balanced, so the physical real-$\gamma$ scan reaches an exact EP rather than a nearest approach.

Starting from the exact scalar equation 
\begin{equation}
\mathcal{F}(E,\gamma)\equiv 1-\frac{i\gamma}{2}G_0(s,s;E)=0,
\label{eq:F_intermediate}
\end{equation}
an exact second-order EP is defined by the double-root conditions 
\begin{equation}
\mathcal{F}(E_{\mathrm{EP}},\gamma_{\mathrm{EP}})=0, \qquad \partial_E 
\mathcal{F}(E_{\mathrm{EP}},\gamma_{\mathrm{EP}})=0.
\label{eq:EP_double_root}
\end{equation}
Since $\partial_E\mathcal{F}=-(i\gamma/2) G_0^{\prime }(s,s;E)$, these
conditions become 
\begin{equation}
G_0^{\prime }(s,s;E_{\mathrm{EP}})=0, \qquad \gamma_{\mathrm{EP}}=\frac{2}{%
iG_0(s,s;E_{\mathrm{EP}})}.  \label{eq:EP_exact_green}
\end{equation}
This form is useful because it shows that an EP is controlled by the local
Green function at the lossy site. The near degeneracy of two clean levels is
not by itself sufficient.

A simple approximation is obtained by retaining only two nearby clean poles
in $G_0(s,s;E)$, 
\begin{equation}
G_0(s,s;E)\approx \frac{w_a}{E-\varepsilon_a}+\frac{w_b}{E-\varepsilon_b},
\qquad w_{a,b}=|\phi_{a,b}(s)|^2.  \label{eq:two_pole_G0}
\end{equation}
Substituting this into Eq.~\eqref{eq:F_intermediate} gives 
\begin{equation}
(E-\varepsilon_a)(E-\varepsilon_b)-\frac{i\gamma}{2}\Big[w_a(E-\varepsilon_b)+w_b(E-%
\varepsilon_a)\Big]=0.  \label{eq:two_pole_quadratic}
\end{equation}
Introducing 
\begin{equation}
\begin{split}
\bar\varepsilon &= \frac{\varepsilon_a+\varepsilon_b}{2}, \quad
\delta = \frac{\varepsilon_b-\varepsilon_a}{2}, \\
W &= w_a+w_b, \quad
\Delta w = w_b-w_a,
\end{split}
\label{eq:W_Deltaw_defs}
\end{equation}
and writing $x=E-\bar\varepsilon$, one obtains 
\begin{equation}
x^2-\delta^2-\frac{i\gamma}{2}\Big[Wx+\delta\,\Delta w\Big]=0,
\label{eq:two_pole_xform}
\end{equation}
with discriminant 
\begin{equation}
\mathcal{D}(\gamma)=4\delta^2-\frac{\gamma^2W^2}{4}+2i\gamma\delta\,\Delta w.
\label{eq:two_pole_discriminant}
\end{equation}

This expression makes the distinction between an exact EP and near-EP
behavior quite clear. For an EP to lie on the real-$\gamma$ axis, one must have $%
\mathcal{D}(\gamma )=0$ for real $\gamma $. Since the discriminant is
complex, both its real and imaginary parts must vanish, 
\begin{equation}
4\delta ^{2}-\frac{\gamma ^{2}W^{2}}{4}=0,\qquad 2\gamma \delta \,\Delta w=0.
\label{eq:real_axis_EP_split}
\end{equation}%
For a nontrivial two-level crossing with $\delta \neq 0$ and $\gamma \neq 0$%
, the second condition requires $w_{a}=w_{b}$. Hence an exact EP on the real 
$\gamma $ axis requires equal local spectral weights on the lossy site. If
this balance is only approximate, the branch point is shifted away from the
real axis. A real parameter scan then shows only a narrow nearest approach.
Such an imbalanced situation gives near-EP behavior rather than an EP. The centered-loss geometry studied below instead realizes the balanced case exactly.

The balanced case is still useful because it gives the local form of the
crossover most clearly. Setting $w_{a}=w_{b}\equiv w$, so that $W=2w$ and $%
\Delta w=0$, Eq.~\eqref{eq:two_pole_xform} yields 
\begin{equation}
E_{\pm }(\gamma )=\bar{\varepsilon}+\frac{i\gamma w}{2}\pm \sqrt{\delta ^{2}-\frac{\gamma
^{2}w^{2}}{4}}.  \label{eq:Epm_balanced}
\end{equation}%
The two branches coalesce at 
\begin{equation}
\gamma _{\mathrm{EP}}=\frac{2|\delta |}{w}=\frac{|\varepsilon
_{b}-\varepsilon _{a}|}{w}.  \label{eq:gamma_EP_balanced}
\end{equation}%
For $\gamma <\gamma _{\mathrm{EP}}$, the square root is real, so the two
branches remain separated in their real parts while sharing the same decay
rate, 
\begin{equation}
\Im E_{+}=\Im E_{-}=\frac{\gamma w}{2}.  \label{eq:equal_decay_below_EP}
\end{equation}%
For $\gamma >\gamma _{\mathrm{EP}}$, the square root becomes purely
imaginary, and the two branches split in their imaginary parts: 
\begin{equation}
\Im E_{\pm }=\frac{\gamma w}{2}\pm \sqrt{\frac{\gamma ^{2}w^{2}}{4}-\delta ^{2}}.
\label{eq:ImE_balanced_above}
\end{equation}%
The lower branch is then 
\begin{equation}
\Im E_{-}(\gamma )\simeq \frac{\delta ^{2}}{\gamma w}\qquad (\gamma \gg
\gamma _{\mathrm{EP}}),  \label{eq:qd_branch_large_gamma}
\end{equation}%
which is the local two-mode form of the $1/\gamma $ suppression that appears
on the strong-loss side.

If this pair lies on the lower envelope of the rapidity spectrum, the same
two-mode approximation gives the local contribution to the Liouvillian gap, 
\begin{equation}
\Delta _{\mathcal{L}}^{(2)}(\gamma )\approx 
\begin{cases}
\gamma w/2, & \gamma \leq \gamma _{\mathrm{EP}}, \\[4pt]
\gamma w/2-\sqrt{\gamma ^{2}w^{2}/4-\delta ^{2}}, & \gamma \geq \gamma _{\mathrm{%
EP}}.%
\end{cases}
\label{eq:gap_two_mode_balanced}
\end{equation}%
At the exact EP one has $\Delta _{\mathcal{L}}^{(2)}=|\delta |$. On the
strong-loss side, the balanced model gives 
\begin{equation}
\Delta _{\mathcal{L}}^{(2)}(\gamma )-\Delta _{\mathcal{L}}^{(2)}(\gamma _{%
\mathrm{EP}})\sim -\sqrt{|\delta |w\,(\gamma -\gamma _{\mathrm{EP}})},
\label{eq:gap_cusp_EP}
\end{equation}%
so the slope is singular at the EP. In a generic finite chain the balance
condition is not exact, and this singularity is rounded into near-EP
behavior. For the symmetry-balanced geometry used below, the singular
coalescence remains on the real-$\gamma$ axis.

The consequence for the many-body spectrum is direct. Every Liouvillian eigenvalue is an additive sum of single-particle rapidities, $\Lambda=\Lambda_{\mathrm{rest}}+\beta_{\pm}$. Therefore, when a rapidity pair reaches an EP, an entire family of many-body branches is rearranged together. For this passive quadratic loss problem, the asymptotic full-space Liouvillian gap satisfies $\Delta_{\mathcal{L}}=\min_j \Im E_j$, as stated in Eq.~\eqref{eq:gap_from_reduction}. If the selected pair belongs to the lower edge of the rapidity spectrum, its reorganization directly modifies the asymptotic gap. If it lies higher in the spectrum, the same spectral rearrangement remains visible in the many-body stripes but does not control the slowest relaxation rate.

To quantify the spectral diagnostics used below, let $E_j(\gamma)$ be the eigenvalues of $P(\gamma)$ and let $\bm\psi_j(\gamma)$ denote the corresponding right eigenvectors. Since $P$ is complex symmetric, the phase rigidity of mode $j$ is defined by
\begin{equation}
\rho_j(\gamma)\equiv \frac{|\bm\psi_j^{T}\bm\psi_j|}{\bm\psi_j^{\dagger}\bm\psi_j},
\label{eq:phase_rigidity_def}
\end{equation}
Let $E_\pm(\gamma)$ and $\bm\psi_\pm(\gamma)$ denote the two continuously tracked rapidities and eigenvectors that form the lower-edge pair in Fig.~\ref{fig:intermediate_gamma}. The pair-resolved spectral distance and phase rigidity plotted in panel (b) are
\begin{align}
d_{\rm pair}(\gamma)
&\equiv |E_+(\gamma)-E_-(\gamma)|,
\label{eq:d_pair_def}\\
\rho_{\rm pair}(\gamma)
&\equiv \min_{\sigma=\pm}\rho_\sigma(\gamma).
\label{eq:rho_pair_def}
\end{align}
In the scan plots below, $\Delta_{\mathcal L}(\gamma)$, $\rho_{\rm pair}(\gamma)$, and $d_{\rm pair}(\gamma)$ are each divided by their respective maxima over the displayed $\gamma$ window so that the gap cusp and the coalescence of this same lower-edge pair can be compared directly.

To connect these spectral quantities to the actual relaxation, we also consider exact one-body dynamics for single-particle initial states. If
\begin{equation}
C_{jk}(t)\equiv \mathrm{Tr}[\rho(t)c_k^{\dagger}c_j]
\label{eq:onebody_C_def}
\end{equation}
is the one-body correlation matrix and the initial state occupies a single site $j_0$, so that $C(0)=|j_0\rangle\langle j_0|$, then the Lindblad evolution remains exactly closed at the one-body level and takes the form
\begin{equation}
\begin{split}
C(t)=U(t)C(0)U^{\dagger}(t),
\qquad
U(t)=e^{Mt},\\
M=-ih-\frac{\Gamma}{2},
\qquad
\Gamma=\gamma |s\rangle\langle s|.
\end{split}
\label{eq:onebody_exact_evolution}
\end{equation}
The local occupation on site $j$ is $n_j(t)=C_{jj}(t)$. Equivalently, in unit-cell notation,
\begin{equation}
n_{n,A}(t)=C_{2n-1,2n-1}(t),
\qquad
n_{n,B}(t)=C_{2n,2n}(t).
\label{eq:local_occ_def}
\end{equation}

The dynamical panels in Fig.~\ref{fig:intermediate_gamma} use these exact one-body observables rather than a projected two-mode approximation (see Appendix~\ref{app:onebody_dynamics} for details). The spectral criterion and the real-space signature can be displayed within the same geometry. We therefore focus on a chain with $N=15$, $t_2=1.0$, and a lossy site at $(m,A)=(8,A)$. Fixing $t_1=0.655$ and scanning the physical Lindblad rate $\gamma$ gives an exact EP on the real-$\gamma$ axis at $\gamma_{\mathrm{EP}}=2.6200$.

Panel~(a) tracks the lower-edge rapidity pair and shows its exact coalescence at $\gamma_{\mathrm{EP}}$. The centered-loss symmetry produces simultaneous coalescences of the other chiral-partner pairs as well; panel~(a) isolates the lower-edge pair because it is the one that controls $\Delta_{\mathcal L}$. Panel~(b) shows that the pair distance and phase rigidity vanish at the same loss rate at which the gap develops its EP cusp. These coincident diagnostics establish genuine defectiveness, rather than a finite avoided crossing.

To connect the exact EP to parity-even one-body dynamics, we distinguish the full-space Liouvillian gap $\Delta_{\mathcal L}$ from the slow one-body decay edge $\Delta_C$. The correlation matrix evolves with exponents $\beta_\mu^{(-)}+\beta_\nu^{(+)}$. Neither parameter set used below contains an exact dark rapidity, and the slowest contributing pair consists of the two symmetry-related lower-edge rapidities; hence $\Delta_C=2\Delta_{\mathcal L}$. We compare the lower-edge EP with a nondefective high-energy spectral control whose $\Delta_C$ is numerically matched to the EP value. We use the lower-edge EP at $(t_1,\gamma)=(0.655,2.6200)$ and a matched high-energy control at $(t_1,\gamma)=(0.8465,2.2421)$. For both points, direct diagonalization gives $\Delta_C\simeq2.005648\times10^{-3}$ and $\Delta_{\mathcal L}\simeq1.002824\times10^{-3}$; the relative mismatch between the two $\Delta_C$ values is approximately $10^{-12}$. The comparison is therefore not driven by different leading one-body exponential rates. At the control point the closest spectral pair lies well above the lower rapidity edge and remains separated, so the lower edge is nondefective. For the initial single-particle excitation on site $(3,A)$, we monitor the exact left-window occupation
\begin{equation}
W_L(t)\equiv \sum_{n=1}^{m}\bigl[n_{n,A}(t)+n_{n,B}(t)\bigr],
\label{eq:WL_def_main}
\end{equation}
and the compensated, anchored signal
\begin{equation}
\widetilde W_L(t)\equiv
\frac{e^{\Delta_C t}W_L(t)}{e^{\Delta_C t_*}W_L(t_*)},
\label{eq:WL_compensated_main}
\end{equation}
where $t_*$ denotes the left end of the displayed late-time window. Panel~(c) of Fig.~\ref{fig:intermediate_gamma} shows the raw decay $W_L(t)/W_L(0)$ for these two points. Because the two one-body decay edges $\Delta_C$ are matched, the more discriminating comparison is provided by panel~(d), which plots $\widetilde W_L(t)$. The lower-edge EP exhibits a clear late-time growth of the compensated signal, reaching a factor of about $5.3$ over the displayed window. By contrast, the matched high-energy control remains nearly flat and ends within about $0.01\%$ of its anchored value. The real-space dynamics therefore confirms the spectral conclusion of panels~(a) and (b): exact defectiveness at the lower rapidity edge produces the polynomially enhanced prefactor, whereas high-energy spectral structure with the same $\Delta_C$ does not reorganize the final asymptotic tail.

The same point can be restated dynamically. Away from an EP, the slow sector is diagonalizable and the late-time signal is a sum of exponentials. At a second-order Jordan block, the one-body propagator contains
\begin{equation}
	e^{Mt}\sim e^{\beta t}(I+tN_J),
\end{equation}
where $N_J$ is nilpotent inside the defective subspace. A single coherent amplitude can therefore acquire at most a linear prefactor. The plotted quantities in Fig.~\ref{fig:intermediate_gamma}(c,d), however, are one-body bilinears governed by $C(t)=e^{Mt}C(0)e^{M^\dagger t}$. Combining the two propagators, a generic bilinear observable may contain a polynomial prefactor up to second order,
\begin{equation}
	\mathcal{O}(t)-\mathcal{O}_{\mathrm{ss}}\sim
	(A_0+A_1t+A_2t^2)e^{-\Delta_C t}.
\end{equation}
The lower-edge parameter set used here realizes this defective case. The matched control remains diagonalizable at the lower rapidity edge, so after compensation by the same $\Delta_C$ its late-time signal stays nearly constant. The short dashed segment in Fig.~\ref{fig:intermediate_gamma}(d) is a local linear fit over a restricted late-time subwindow and is included only as a visual polynomial guide; it is not used to assign the global polynomial degree. We therefore use the compensated trace only as evidence for a polynomially enhanced late-time prefactor, without assigning a fixed polynomial order over the full plotted window. This directly separates defectiveness at the lower edge from high-energy spectral structure that does not govern the asymptotic tail. The intermediate-loss regime therefore provides the crossover between the linear weak-loss opening of decay channels and the $1/\gamma$ suppression derived in the next section.

\section{Large loss regime: Zeno splitting and geometry of the slow sector}

\label{sec:zeno}

We finally turn to the opposite limit $\gamma\to\infty$ for a single dissipative defect at site $s$. The use of ``Zeno'' here refers to a separation between a rapidly decaying local subspace and an active slow subspace, rather than to a complete freezing of the chain. An amplitude on the lossy site relaxes at rate $\gamma/2$, so for $\gamma\gg |t_1|,|t_2|$ that site follows the neighboring amplitudes only virtually and can be adiabatically eliminated. To leading order, the slow spectrum is therefore that of the two open subchains obtained by removing site $s$; at finite $\gamma$, the slow subspace visits that site only virtually, producing corrections of order $O(J^2/\gamma)$, where $J$ denotes a representative hopping scale. The counterintuitive decrease of the induced slow decay as the local loss is increased, together with the projection of the lossy site out of the slow dynamics, is the dissipative Zeno effect in this model~\cite{FacchiPascazio2002,Syassen2008,Barontini2013}. The Liouvillian gap is consequently determined by the spectra and boundary amplitudes of the two reduced subchains.

We start from the exact secular equation derived in Sec.~\ref%
{sec:exact_rapidity}, 
\begin{equation}
D_{2N}(E)-\frac{i\gamma}{2}\,D_{s-1}(E)\,D_{2N-s}(E)=0,  \label{eq:exact_secular_zeno}
\end{equation}%
where $D_{\ell }(E)=\det (E\boldsymbol{I}_{\ell }-P_{0,\ell })$ is the clean
tridiagonal determinant of a segment of length $\ell $. Dividing Eq.~%
\eqref{eq:exact_secular_zeno} by $\gamma $ and taking $\gamma \rightarrow
\infty $ with $E=O(1)$ fixed, we obtain 
\begin{equation}
D_{s-1}(E)D_{2N-s}(E)=0.  \label{eq:zeno_factorization}
\end{equation}%
Hence the $O(1)$ spectrum in the Zeno limit is 
\begin{eqnarray}
\{E\}_{\gamma \rightarrow \infty } &=&\{E_{\mu }^{(L)}:\ D_{s-1}(E_{\mu
}^{(L)})=0\}  \notag \\
\cup \{E_{\nu }^{(R)} &:&\ D_{2N-s}(E_{\nu }^{(R)})=0\}.
\label{eq:spectrum_union}
\end{eqnarray}%
The original chain is therefore reduced, at the level of the slow spectrum,
to a left open segment of length $L_{L}=s-1$ and a right open segment of
length $L_{R}=2N-s$. Changing the defect position changes both $(L_{L},L_{R})
$ and the termination of the two segments, and this determines the slow
decay scale.

Besides the $O(1)$ roots in Eq.~\eqref{eq:spectrum_union}, Eq.~%
\eqref{eq:exact_secular_zeno} also admits one root with $|E|\sim \gamma $.
For $|E|\gg |t_{1}|,|t_{2}|$, 
\begin{eqnarray}
D_{2N}(E) &\sim &E^{2N},\qquad  \\
D_{s-1}(E)D_{2N-s}(E) &\sim &E^{2N-1},
\end{eqnarray}%
so Eq.~\eqref{eq:exact_secular_zeno} gives 
\begin{equation}
E^{2N}-\frac{i\gamma}{2} E^{2N-1}\approx 0\quad \Longrightarrow \quad E_{Z}\approx
\frac{i\gamma}{2}.  \label{eq:fast_mode}
\end{equation}%
This root corresponds to a defect-localized ultrafast mode. In the rapidity
plane it becomes $\beta _{Z}\approx -\gamma/2$, well separated from the
slow part of the spectrum.

We next derive the leading correction for large but finite $\gamma $. Split
the Hilbert space into the lossy site $s$ and the remaining subspace $R$. In
the basis $\{|s\rangle \}\oplus R$, the matrix $P$ takes the block form 
\begin{equation}
P=%
\begin{pmatrix}
i\gamma/2  & \bm v^{T} \\ 
\bm v & P_{R}%
\end{pmatrix}%
,  \label{eq:blockP}
\end{equation}%
where $P_{R}$ is the restriction of $P_{0}$ to the reduced space $R$, and $%
\bm v$ contains the hoppings from site $s$ to its neighbors in $R$. For the
tridiagonal SSH chain, site $s$ has at most two neighbors, so $\bm v$ has at
most two nonzero components. Then, for a bulk defect with $1<s<2N$, 
\begin{equation}
\bm v=J_{s-1}\bm e_{s-1}^{(R)}+J_{s}\bm e_{s+1}^{(R)}.
\label{eq:v_general_bulk}
\end{equation}%
At a boundary, one of these two terms is absent.

Let $P\bm\psi=E\bm\psi$ with $\bm\psi=(\psi_s,\bm\psi_R)^T$. The component
on the lossy site satisfies 
\begin{equation}
\frac{i\gamma}{2}\psi_s+\bm v^T\bm\psi_R=E\psi_s,
\end{equation}
so that 
\begin{equation}
\psi_s=-\frac{\bm v^T\bm\psi_R}{i\gamma/2-E}.  \label{eq:psi_s_elim_P}
\end{equation}
Substituting this into the equation for $\bm\psi_R$ gives the exact reduced
eigenvalue problem 
\begin{equation}
\left(P_R-\frac{\bm v\bm v^T}{i\gamma/2-E}\right)\bm\psi_R=E\bm\psi_R.
\label{eq:exact_reduced_P}
\end{equation}

We are interested in the slow sector with $E=O(1)$ as $\gamma\to\infty$. In
this regime, 
\begin{equation}
\frac{1}{i\gamma/2-E} = \frac{2}{i\gamma}\frac{1}{1-2E/(i\gamma)} = \frac{2}{i\gamma}+O(\gamma^{-2}),  \label{eq:denom_expand_P}
\end{equation}
and Eq.~\eqref{eq:exact_reduced_P} becomes 
\begin{equation}
\left(P_R+\frac{2i}{\gamma}\bm v\bm v^T\right)\bm\psi_R = E\bm\psi_R \qquad %
\bigl(\text{errors }O(\gamma^{-2})\bigr).  \label{eq:PZ_def}
\end{equation}
The same elimination makes the Zeno suppression directly visible in the lossy-site amplitude. Combining Eqs.~\eqref{eq:psi_s_elim_P} and \eqref{eq:denom_expand_P} gives
\begin{equation}
\psi_s=\frac{2i}{\gamma}\,\bm v^T\bm\psi_R+O(\gamma^{-2}).
\label{eq:psi_s_zeno_scaling}
\end{equation}
For a normalized one-particle mode, the occupation of the lossy site and the associated instantaneous loss current therefore scale as
\begin{align}
|\psi_s|^2
&=\frac{4}{\gamma^2}|\bm v^T\bm\psi_R|^2+O(\gamma^{-3}),
\label{eq:lossy_occupation_zeno}\\
\mathcal J_{\rm loss}\equiv\gamma|\psi_s|^2
&=\frac{4}{\gamma}|\bm v^T\bm\psi_R|^2+O(\gamma^{-2}).
\label{eq:loss_current_zeno}
\end{align}
Thus stronger bare loss suppresses the occupation of the monitored site as $\gamma^{-2}$ and reduces the residual leakage from the slow subspace as $\gamma^{-1}$, rather than freezing the two surviving subchains.
Thus the slow sector is governed by the effective matrix 
\begin{equation}
P_Z \equiv P_R+\frac{2i}{\gamma}\bm v\bm v^T.  \label{eq:PZ_compact}
\end{equation}
The Zeno reduction therefore maps the problem to a weak imaginary rank-one
perturbation on the reduced chain, now with effective strength $2/\gamma$. Its diagonal terms describe the residual loss inherited by modes adjacent to the removed site, while its off-diagonal terms mediate cross-defect transfer of order $J_LJ_R/\gamma$. Both effects vanish as the bare loss grows. This is the explicit Schur-complement realization of the intuitive Zeno projection described above.

This already shows that the slow eigenvalues satisfy $\Im E=O(1/\gamma)$, or
equivalently $-\Re\beta=O(1/\gamma)$. To obtain the prefactor, let $P_R\bm%
\psi_m=E_m^{(0)}\bm\psi_m$ be a normalized eigenpair of the reduced
Hermitian chain. Repeating the weak-perturbation argument used in Sec.~\ref%
{sec:weak_gamma}, one finds 
\begin{equation}
E_m = E_m^{(0)}+\frac{2i|\bm v^T\bm\psi_m|^2}{\gamma}+O(\gamma^{-2}),
\label{eq:ImE_zeno_general}
\end{equation}
and hence 
\begin{equation}
-\Re\beta_m = \Im E_m = \frac{2|\bm v^T\bm\psi_m|^2}{\gamma}+O(\gamma^{-2}).
\label{eq:decay_zeno_general}
\end{equation}
Taking the smallest nonzero decay rate gives 
\begin{equation}
\Delta_{\mathcal{L}} \asymp \frac{2}{\gamma}\min_m |\bm v^T\bm\psi_m|^2
\qquad (\gamma\gg |t_1|,|t_2|),  \label{eq:gap_zeno_general}
\end{equation}
where $\{\bm\psi_m\}$ are the eigenmodes of $P_R$. If a reduced-chain eigenvalue is degenerate, the rank-one correction $\bm v\bm v^T$ must first be diagonalized within that degenerate subspace before the smallest nonzero decay rate is selected.

Since removing the lossy site splits the system into two open subchains,
each eigenmode is supported predominantly on either the left or the right
segment. For a left-subchain mode, 
\begin{equation}
\bm v^T\bm\psi_m \approx J_{s-1}\psi_m(s-1),  \label{eq:v_overlap_left}
\end{equation}
whereas for a right-subchain mode, 
\begin{equation}
\bm v^T\bm\psi_m \approx J_s\psi_m(s+1).  \label{eq:v_overlap_right}
\end{equation}
The Zeno gap is therefore controlled by the wave-function weights at the two
sites adjacent to the removed defect.

\subsection{Gap scaling in the Zeno regime}

\label{subsec:zeno_gap_scaling}

We first consider the generic case in which the slowest modes are extended
standing waves on the two open segments. For an open chain of length $L$,
the quantized momenta scale as 
\begin{equation}
k_m\sim \frac{m\pi}{L+1}, \qquad m=1,2,\dots,  \label{eq:k_quant_zeno}
\end{equation}
so $k_{\min}\sim \pi/(L+1)\sim L^{-1}$. Near an open boundary, the
standing-wave envelope behaves as $\psi_k(n)\propto \sin(nk)$, hence $%
\psi_k(1)\propto \sin k\sim k$ for $k\to 0$. After normalization over $O(L)$
sites, one has $|\psi_k|\sim L^{-1/2}$, and therefore 
\begin{equation}
|\psi_{k_{\min}}(1)|^2 \sim \left(\frac{\sin k_{\min}}{\sqrt{L}}\right)^2
\sim \frac{k_{\min}^2}{L} \sim \frac{1}{L^3}.
\label{eq:boundary_weight_Lm3_zeno}
\end{equation}

Applying Eq.~\eqref{eq:gap_zeno_general} to the left segment of length $%
L_L=s-1$ gives 
\begin{equation}
\Delta_{\mathcal{L}}^{(L)} \asymp \frac{2J_{s-1}^2}{\gamma}\frac{1}{L_L^3},
\label{eq:gap_zeno_left_poly}
\end{equation}
while for the right segment of length $L_R=2N-s$, 
\begin{equation}
\Delta_{\mathcal{L}}^{(R)} \asymp \frac{2J_s^2}{\gamma}\frac{1}{L_R^3}.
\label{eq:gap_zeno_right_poly}
\end{equation}
Hence the overall gap is 
\begin{equation}
\begin{split}
\Delta_{\mathcal{L}} \asymp \frac{2}{\gamma} \min\left\{ \frac{J_{s-1}^2}{%
(s-1)^3}, \frac{J_s^2}{(2N-s)^3} \right\} \\
 (\text{generic slow sector}%
).  
\end{split}
\label{eq:gap_zeno_poly_general_s}
\end{equation}
For $s=O(N)$ and $J_{s-1},J_s=O(t_1,t_2)$, this reduces to $\Delta_{\mathcal{%
L}}\asymp (2t^2/\gamma)N^{-3}$ up to an $O(1)$ geometric prefactor.

In the topological phase $|t_1|<|t_2|$, the cut segments may support
edge-localized modes. If the termination at the cut makes the coupling to
the rank-one channel $\bm v\bm v^T$ vanish at leading order, the overlap $|%
\bm v^T\bm\psi|$ becomes parametrically small, and the Zeno gap is then
exponentially suppressed.

To see this explicitly, consider a termination for which the removed site is
a $B$ site coupled to two neighboring $A$ amplitudes $a_1=\psi(1A)$ and $%
a_2=\psi(2A)$. In that case, 
\begin{equation}
\bm v^T\bm\psi \propto t_1 a_1+t_2 a_2.  \label{eq:vTpsi_local_AA}
\end{equation}
For a topological left-edge mode, the clean zero-energy recursion gives 
\begin{equation}
t_1 a_1+t_2 a_2=0,  \label{eq:edge_dark_constraint_zeno}
\end{equation}
so that 
\begin{equation}
\bm v^T\bm\psi_{\mathrm{edge}}=0 \qquad (\text{thermodynamic limit}).
\label{eq:v_orth_edge_zeno}
\end{equation}
Thus the edge mode is decoupled from the effective Zeno loss channel at
leading order.

For a finite segment, the two edge modes hybridize through their
exponentially small overlap. Writing 
\begin{equation}
r=\left|\frac{t_1}{t_2}\right|<1,  \label{eq:r_zeno}
\end{equation}
the violation of Eq.~\eqref{eq:edge_dark_constraint_zeno} is of order $r^{N_{%
\mathrm{eff}}}$, where $N_{\mathrm{eff}}$ is the effective unit-cell length
of the relevant segment. Equivalently, 
\begin{equation}
|\bm v^T\bm\psi_{\mathrm{edge}}|\sim r^{N_{\mathrm{eff}}}, \qquad |\bm v^T\bm%
\psi_{\mathrm{edge}}|^2\sim r^{2N_{\mathrm{eff}}}.
\label{eq:v_overlap_edge_finiteN_zeno}
\end{equation}
Substituting this into Eq.~\eqref{eq:gap_zeno_general}, we obtain 
\begin{equation}
\begin{split}
&\Delta_{\mathcal{L}} \asymp \frac{2t^2}{\gamma}r^{2N_{\mathrm{eff}}} \\
&(\gamma\gg |t_1|,|t_2|,\ |t_1|<|t_2|,\ \text{dark termination}).
\end{split}
\label{eq:gap_zeno_exp}
\end{equation}
Compared with the weak-loss result $\Delta_{\mathcal{L}}\asymp (\gamma/2) r^{2N_{\mathrm{eff}}}$, the exponential finite-size factor is unchanged, while the
overall prefactor is replaced by $2t^2/\gamma$.

The strong-loss spectrum therefore has a simple structure. Most rapidities form two slow groups that approach the spectra of the two open segments and whose decay scales are controlled by Eq.~\eqref{eq:gap_zeno_general}. In addition, one isolated rapidity moves to $|\beta|\sim\gamma/2$, corresponding to the defect-localized fast mode in Eq.~\eqref{eq:fast_mode}. The position of the defect enters only through the lengths and terminations of the two reduced segments, and these determine whether the Zeno gap follows the algebraic form in Eq.~\eqref{eq:gap_zeno_poly_general_s} or the exponentially small form in Eq.~\eqref{eq:gap_zeno_exp}. This closes the connection between the weak-loss selection rule, the intermediate reorganization window, and the geometry-controlled Zeno limit.

\begin{figure*}[t]
\centering
\includegraphics[width=\textwidth]{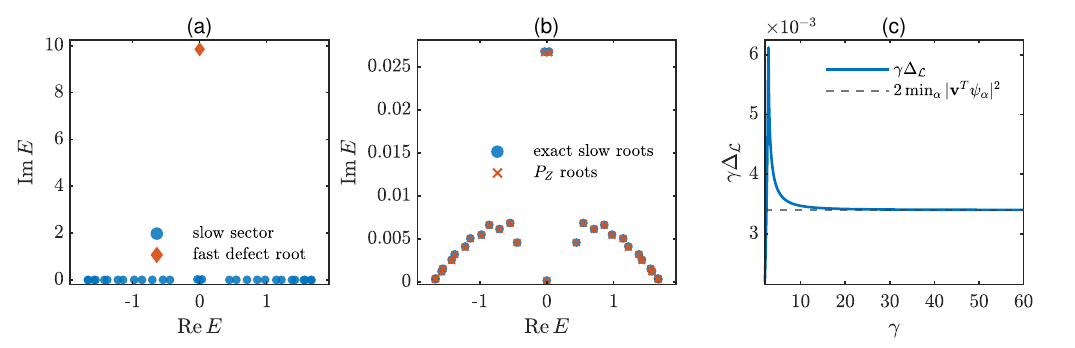}%
\caption{Strong-loss Zeno regime for a chain with $N=15$, $t_1=0.70$, $t_2=1.0$, and loss on $(m,A)=(8,A)$. Panels (a) and (b) use the representative value $\gamma=20$. (a) Exact rapidity spectrum. One ultrafast defect root is clearly separated from the remaining slow sector, in agreement with Eq.~\eqref{eq:fast_mode}. (b) Comparison between the exact slow rapidities and the roots of the effective Zeno matrix $P_Z$ defined in Eq.~\eqref{eq:PZ_for_figure}. The two sets of points nearly coincide, showing that the slow sector is governed by the reduced chain with the lossy site removed. (c) Zeno scaling of the Liouvillian gap for the same geometry. The numerical data for $\gamma\Delta_{\mathcal L}$ approach a large-$\gamma$ plateau, which is equivalent to the asymptotic law $\Delta_{\mathcal L}\propto 1/\gamma$ predicted by Eq.~\eqref{eq:zeno_scaling_for_figure}. The dashed line is the plateau expected from the corrected reduced-chain overlap factor $2\min_m |\bm v^T \bm\psi_m|^2$.}
\label{fig:large_gamma}
\end{figure*}

Figure~\ref{fig:large_gamma} summarizes the strong-loss regime using the
three quantities directly tied to the analytical Zeno reduction. For the numerical illustration we fix $N=15$, $t_1=0.70$, $t_2=1.0$, and place the loss on the bulk site $(m,A)=(8,A)$. Panels~(a) and (b) use the representative large-loss value $\gamma=20$, while panel~(c) scans $\gamma$ over the Zeno window. Panel~(a)
shows the full complex rapidity spectrum at this representative large loss. One
root is separated far above the rest along the imaginary direction,
corresponding to the ultrafast defect mode of Eq.~\eqref{eq:fast_mode},
while the remaining roots form a slow sector with $\Im E=O(1/\gamma)$.

Panel~(b) compares this exact slow sector with the roots of the effective
Zeno matrix
\begin{equation}
P_Z=P_R+\frac{2i}{\gamma}\bm v\bm v^T,
\label{eq:PZ_for_figure}
\end{equation}
introduced in Eq.~\eqref{eq:PZ_compact}. The close agreement confirms that
the long-time Liouvillian sector is governed by the cut-chain problem rather
than by the original defect site.

Panel~(c) tests the scaling law
\begin{equation}
\Delta_{\mathcal L}\asymp \frac{2}{\gamma}\min_m |\bm v^T\bm\psi_m|^2,
\label{eq:zeno_scaling_for_figure}
\end{equation}
derived in Eq.~\eqref{eq:gap_zeno_general}. Equivalently, the product
$\gamma\Delta_{\mathcal L}$ should approach a constant equal to twice the reduced-chain overlap factor for large $\gamma$.
The plateau seen numerically is therefore the direct signature of the Zeno
suppression of the slow decay scale.

The numerical data in Fig.~\ref{fig:large_gamma} thus support the three
analytical ingredients of the strong-loss regime: separation of one fast
defect root, accurate reproduction of the slow sector by the reduced Zeno
matrix $P_Z$, and the asymptotic $1/\gamma$ law for the Liouvillian gap.

\section{Conclusion}

A single bulk loss in an open SSH chain provides an exactly solvable setting in which the Liouvillian gap can be followed continuously from weak loss to the Zeno limit within the original many-body Lindblad problem. The key simplification is an exact reduction to a finite non-Hermitian single-particle matrix with a rank-one imaginary impurity. Its rapidities generate both operator-parity sectors of the Liouvillian spectrum: the smallest positive rapidity decay fixes the full-space gap, while parity-even observables relax through the corresponding even subset sums.

This reduction exposes three regimes. In the weak-loss regime, the local spectral weight at the lossy site selects the clean modes that first acquire decay, giving the generic law $\Delta_{\mathcal{L}}\sim \gamma N^{-3}$ and, in the topological phase, exponentially smaller edge-controlled gaps. In the intermediate regime, the exact secular equation shows that an exceptional point on the real-$\gamma$ axis requires balanced local weights. The centered-loss geometry enforces this balance and produces simultaneous symmetry-related coalescences, including an exact EP at the lower rapidity edge. Comparison with a nondefective high-energy control having the same parity-even decay edge shows that only lower-edge defectiveness generates the polynomially enhanced late-time prefactor. In the strong-loss regime, one ultrafast defect mode separates from an active slow sector in which the lossy site has been projected out. Transfer across the defect and the residual induced decay are virtual $O(J^2/\gamma)$ processes, so increasing the bare loss slows the remaining relaxation and yields $\Delta_{\mathcal L}\propto\gamma^{-1}$. This local projection, rather than a literal freezing of all degrees of freedom, is the dissipative Zeno effect established by the Schur-complement reduction.

The disorder analysis in Appendix~\ref{app:disorder} further separates robust mechanisms from parameter-sensitive spectral details. For bond disorder that preserves SSH chiral symmetry, the ensemble-median gap retains the weak-loss law $\Delta_{\mathcal L}\propto\gamma$ and the strong-loss Zeno law $\Delta_{\mathcal L}\propto\gamma^{-1}$. The finite-size $N^{-3}$ trend also remains visible at genuinely weak disorder. Within the finite chains and disorder range sampled here, exact imaginary-axis EPs are found realization by realization, but their positions broaden and the probability that the lowest exact EP still controls the rapidity edge decreases with disorder. Thus the weak- and strong-loss mechanisms are structurally robust, whereas the lower-edge placement of an intermediate-loss EP is more sensitive to geometry and spectral ordering.

The original formulation is a fermionic many-body problem, and the finite $4^{N_s}$ Liouvillian spectrum, operator-parity sectors, and fermionic subset sums are genuinely statistics dependent. The distinct rapidity values and the weak-loss, exceptional-point, and Zeno mechanisms, however, are determined entirely by the linear drift matrix. They can therefore be tested in noninteracting fermionic cold atoms through site-resolved densities and window occupations, in superconducting or photonic resonator arrays through complex amplitudes, resonance linewidths, transmission, and intensity decay, and in classical coupled-mode networks through the corresponding field dynamics. Bosonic and classical realizations reproduce the rapidity problem and its one-body observables, but not the finite fermionic parity-sector organization of the complete Liouvillian. This separation identifies precisely which conclusions are universal to linear open lattices and which retain specifically fermionic many-body content.

\section*{ACKNOWLEDGMENTS}

We acknowledge the support of the National Natural Science Foundation of China (Grants No. 12275193, 11975166).

\appendix

\section{Exact third-quantization reduction}
\label{app:third_quantization}

This Appendix derives the operator-space reduction used in the main text. We
keep the physical Lindblad rate, the operator-parity sectors, and every factor
of two explicit. The derivation follows the fermionic operator-space method of
Refs.~\cite{Prosen2008,Prosen2010}, but all conventions are reconstructed from
the master equation and independently checked against direct Liouville-space
matrices.

\subsection{Lindblad convention, conditional amplitudes, and even bilinears}

The master equation is
\begin{equation}
\dot\rho=-i[H,\rho]+\gamma c_s\rho c_s^\dagger
-\frac{\gamma}{2}\{c_s^\dagger c_s,\rho\},
\label{eq:appA_lindblad_start}
\end{equation}
with
\begin{equation}
H=\bm c^\dagger h\bm c,\qquad
\Gamma=\gamma\Pi_s,\qquad
\Pi_s=|s\rangle\langle s|,
\label{eq:appA_h_Gamma_def}
\end{equation}
where \(\bm c=(c_1,\ldots,c_{N_s})^T\) and \(N_s=2N\). Since
\begin{equation}
L^\dagger L=\gamma c_s^\dagger c_s=\bm c^\dagger\Gamma\bm c,
\label{eq:appA_LdagL}
\end{equation}
the no-jump Hamiltonian is
\begin{equation}
H_{\rm eff}=H-\frac{i}{2}L^\dagger L
=\bm c^\dagger h_{\rm eff}\bm c,
\qquad
h_{\rm eff}=h-\frac{i}{2}\Gamma .
\label{eq:appA_Heff}
\end{equation}
Equation~\eqref{eq:appA_lindblad_start} can therefore be written as
\begin{equation}
\mathcal L[\rho]
=-i\left(H_{\rm eff}\rho-\rho H_{\rm eff}^\dagger\right)
+\gamma c_s\rho c_s^\dagger .
\label{eq:appA_nojump_form}
\end{equation}
The first term is the conditional no-jump evolution, whereas the second term is
the recycling process after a loss event.

The amplitude-loss factor follows directly in the one-particle sector. If a
conditional state \(|\psi(t)\rangle=\sum_j\psi_j(t)|j\rangle\) obeys
\begin{equation}
i\partial_t|\psi(t)\rangle=h_{\rm eff}|\psi(t)\rangle,
\label{eq:appA_conditional_schrodinger}
\end{equation}
then its lossy-site component satisfies
\begin{equation}
\dot\psi_s=-i\sum_k h_{sk}\psi_k-\frac{\gamma}{2}\psi_s.
\label{eq:appA_amplitude_decay}
\end{equation}
Thus \(\gamma/2\) is the conditional amplitude-decay rate. By contrast, the
dissipative contribution to the local population \(n_s=c_s^\dagger c_s\) is
\begin{equation}
\left.\mathcal L^\dagger[n_s]\right|_{\rm dis}
=\gamma c_s^\dagger n_s c_s-\frac{\gamma}{2}\{n_s,n_s\}
=-\gamma n_s,
\label{eq:appA_population_decay}
\end{equation}
because \(c_s^\dagger n_s c_s=0\) and \(n_s^2=n_s\). This establishes the
physical distinction between the population-loss rate \(\gamma\) and the
amplitude-loss rate \(\gamma/2\) without invoking a parity-odd Heisenberg
operator identity.

The exact closure relevant to physical one-body observables is obtained from
the even bilinear correlation matrix
\begin{equation}
C_{jk}(t)=\operatorname{Tr}[\rho(t)c_k^\dagger c_j].
\label{eq:appA_C_def}
\end{equation}
Using the canonical anticommutation relations in
\(\mathcal L^\dagger[c_k^\dagger c_j]\), the Hamiltonian and dissipative
contributions are
\begin{align}
\left.\dot C_{jk}\right|_H
&=-i\sum_\ell h_{j\ell}C_{\ell k}
+i\sum_\ell C_{j\ell}h_{\ell k},
\label{eq:appA_C_H_component}\\
\left.\dot C_{jk}\right|_{\rm dis}
&=-\frac{\gamma}{2}\delta_{js}C_{sk}
-\frac{\gamma}{2}\delta_{ks}C_{js}.
\label{eq:appA_C_dis_component}
\end{align}
Consequently,
\begin{equation}
\begin{aligned}
\dot C&=-i[h,C]-\frac12\{\Gamma,C\}=MC+CM^\dagger,\\
M&=-ih-\frac{\Gamma}{2}.
\end{aligned}
\label{eq:appA_C_matrix}
\end{equation}
Appendix~\ref{app:onebody_dynamics} gives the same bilinear calculation in the
notation used for the real-time figures.

\subsection{Liouville space, parity, and canonical superfermions}

The physical Fock space has dimension \(2^{N_s}\). Its operator space
\(\mathcal K\) has dimension \(4^{N_s}\) and is equipped with the
Hilbert--Schmidt product
\begin{equation}
\langle\!\langle A|B\rangle\!\rangle
=\operatorname{Tr}(A^\dagger B).
\label{eq:appA_HS_inner}
\end{equation}
We vectorize an operator as \(|A\rangle\!\rangle\). Left and right
multiplication are defined by
\begin{align}
\hat c_j^{L}|A\rangle\!\rangle&=|c_jA\rangle\!\rangle,
&\hat c_j^{R}|A\rangle\!\rangle&=|Ac_j\rangle\!\rangle,
\label{eq:appA_left_right_ann}\\
\hat c_j^{\dagger L}|A\rangle\!\rangle&=|c_j^\dagger A\rangle\!\rangle,
&\hat c_j^{\dagger R}|A\rangle\!\rangle&=|Ac_j^\dagger\rangle\!\rangle.
\label{eq:appA_left_right_cre}
\end{align}
Raw left and right maps commute as linear maps, while multiplication on the
right reverses operator order, for example
\begin{equation}
\hat c_j^{\dagger R}\hat c_k^R
=\widehat{(c_kc_j^\dagger)}^{R}.
\label{eq:appA_right_reverse}
\end{equation}

Let
\begin{equation}
\mathcal P=(-1)^{\hat N},\qquad
\hat N=\sum_j c_j^\dagger c_j,
\label{eq:appA_physical_parity}
\end{equation}
and define the operator-parity superoperator by
\begin{equation}
\hat{\mathcal P}|A\rangle\!\rangle
=|\mathcal P A\mathcal P\rangle\!\rangle.
\label{eq:appA_parity_super}
\end{equation}
Every term in Eq.~\eqref{eq:appA_lindblad_start} is parity even, so
\([\widehat{\mathcal L},\hat{\mathcal P}]=0\). The operator space therefore
splits into invariant sectors
\begin{equation}
\mathcal K=\mathcal K_+\oplus\mathcal K_-,\qquad
\hat{\mathcal P}|A_p\rangle\!\rangle=p|A_p\rangle\!\rangle,
\qquad p=\pm1,
\label{eq:appA_parity_sectors}
\end{equation}
with \(\dim\mathcal K_+=\dim\mathcal K_-=2^{2N_s-1}\).

The parity superoperator anticommutes with every odd left or right
multiplication map. We define
\begin{equation}
\hat a_j=\hat c_j^L,
\qquad
\hat a_j'=\hat c_j^{\dagger L},
\qquad
\hat b_j=-\hat{\mathcal P}\hat c_j^{\dagger R},
\qquad
\hat b_j'=\hat{\mathcal P}\hat c_j^R.
\label{eq:appA_superfermion_defs}
\end{equation}
The prime labels an almost-canonical creation map in Liouville space; it is
not an assertion of Hermitian adjunction. The left maps immediately obey
\begin{equation}
\{\hat a_j,\hat a_k\}=\{\hat a_j',\hat a_k'\}=0,
\qquad
\{\hat a_j,\hat a_k'\}=\delta_{jk}.
\label{eq:appA_a_CAR}
\end{equation}
For the right maps,
\begin{align}
\hat b_j\hat b_k'
&=\hat c_j^{\dagger R}\hat c_k^R
=\widehat{(c_kc_j^\dagger)}^R,
\nonumber\\
\hat b_k'\hat b_j
&=\hat c_k^R\hat c_j^{\dagger R}
=\widehat{(c_j^\dagger c_k)}^R,
\label{eq:appA_b_product}
\end{align}
so that
\begin{equation}
\{\hat b_j,\hat b_k'\}=\delta_{jk},
\qquad
\{\hat b_j,\hat b_k\}=\{\hat b_j',\hat b_k'\}=0.
\label{eq:appA_b_CAR}
\end{equation}
Because raw left and right maps commute while \(\hat{\mathcal P}\)
anticommutes with each odd map, every mixed anticommutator between the
\(a\)- and \(b\)-families vanishes. Thus the \(2N_s\) annihilation maps
\((\hat{\bm a},\hat{\bm b})\) and the \(2N_s\) primed maps
\((\hat{\bm a}',\hat{\bm b}')\) form a canonical operator-space fermion
algebra.

\subsection{Parity-resolved quadratic Liouvillian and structure matrices}

We now translate every term of Eq.~\eqref{eq:appA_lindblad_start}. The
Hamiltonian and anticommutator terms give
\begin{align}
-iH\rho&\longrightarrow-i\sum_{jk}h_{jk}\hat a_j'\hat a_k,
\label{eq:appA_H_left}\\
i\rho H&\longrightarrow i\sum_{jk}h_{jk}\hat b_k'\hat b_j,
\label{eq:appA_H_right}\\
-\frac12(\bm c^\dagger\Gamma\bm c)\rho
&\longrightarrow-\frac12\sum_{jk}\Gamma_{jk}\hat a_j'\hat a_k,
\label{eq:appA_loss_left}\\
-\frac12\rho(\bm c^\dagger\Gamma\bm c)
&\longrightarrow-\frac12\sum_{jk}\Gamma_{jk}\hat b_k'\hat b_j.
\label{eq:appA_loss_right}
\end{align}
The recycling sign must be evaluated separately in each parity sector. If
\(|A_p\rangle\!\rangle\in\mathcal K_p\), then
\(|A_pc_k^\dagger\rangle\!\rangle\) has parity \(-p\), and therefore
\begin{equation}
\hat b_k|A_p\rangle\!\rangle
=-\hat{\mathcal P}|A_pc_k^\dagger\rangle\!\rangle
=p|A_pc_k^\dagger\rangle\!\rangle.
\label{eq:appA_b_on_sector}
\end{equation}
It follows that
\begin{equation}
c_jA_pc_k^\dagger
\quad\longrightarrow\quad
p\,\hat a_j\hat b_k|A_p\rangle\!\rangle.
\label{eq:appA_recycling_sector}
\end{equation}
Hence the restriction of the Liouvillian to \(\mathcal K_p\) is
\begin{equation}
\widehat{\mathcal L}_p
=\hat{\bm a}'^{\,T}M\hat{\bm a}
+\hat{\bm b}'^{\,T}M^*\hat{\bm b}
+p\,\hat{\bm a}^{T}\Gamma\hat{\bm b},
\qquad p=\pm1,
\label{eq:appA_Lp_ab}
\end{equation}
Here \(M=-ih-\Gamma/2\) is defined in Eq.~\eqref{eq:appA_C_matrix}. This sector-dependent
sign is essential: the plus sign is correct in the even sector and the minus
sign in the odd sector.

Define
\begin{equation}
\hat{\bm f}=\begin{pmatrix}\hat{\bm a}\\ \hat{\bm b}\end{pmatrix},
\qquad
\hat{\bm f}'=\begin{pmatrix}\hat{\bm a}'\\ \hat{\bm b}'\end{pmatrix}.
\label{eq:appA_f_vectors}
\end{equation}
Then
\begin{equation}
\widehat{\mathcal L}_p
=\hat{\bm f}'^{\,T}\mathbb X\hat{\bm f}
+\frac12\hat{\bm f}^{T}\mathbb Y_p\hat{\bm f},
\label{eq:appA_XY_form}
\end{equation}
with the explicit \(2N_s\times2N_s\) matrices
\begin{equation}
\mathbb X=
\begin{pmatrix}M&0\\0&M^*\end{pmatrix},
\qquad
\mathbb Y_p=
\begin{pmatrix}0&p\Gamma\\-p\Gamma^T&0\end{pmatrix}.
\label{eq:appA_XY_matrices}
\end{equation}
The factor \(1/2\) is fixed by antisymmetry:
\begin{equation}
\frac12\hat{\bm f}^{T}\mathbb Y_p\hat{\bm f}
=p\,\hat{\bm a}^{T}\Gamma\hat{\bm b}.
\label{eq:appA_Y_factor}
\end{equation}
The parity dressing in Eq.~\eqref{eq:appA_superfermion_defs} is an
operator-valued definition map, not an ordinary numerical similarity matrix.
Once a vectorization convention is chosen, every map in
Eq.~\eqref{eq:appA_superfermion_defs} becomes an ordinary
\(4^{N_s}\times4^{N_s}\) matrix; this explicit representation is used in the
numerical checks below.

The number operator
\begin{equation}
\widehat N_f=\hat{\bm f}'^{\,T}\hat{\bm f}
\label{eq:appA_super_number}
\end{equation}
is preserved by the \(\mathbb X\) term, whereas the \(\mathbb Y_p\) term
lowers it by two. Consequently, in a basis ordered by decreasing \(N_f\),
\(\widehat{\mathcal L}_p\) is block lower triangular. The even sector contains
even \(N_f\), the odd sector contains odd \(N_f\), and both sectors have the
same diagonal one-body matrix \(\mathbb X\). The recycling block changes the
eigenoperators but does not change the elementary rapidity eigenvalues.

\subsection{Biorthogonal normal master modes}

Away from an exceptional point, suppose \(\mathbb X\) is diagonalizable. Let
\begin{equation}
\begin{aligned}
\mathbb X R&=RB,
&V\mathbb X&=BV,\\
VR&=I,
&B&=\operatorname{diag}(\beta_1,\ldots,\beta_{2N_s}).
\end{aligned}
\label{eq:appA_X_biorth}
\end{equation}
Define
\begin{equation}
\hat{\bm d}=V\hat{\bm f},
\qquad
\hat{\bm d}'^{\,T}=\hat{\bm f}'^{\,T}R.
\label{eq:appA_d_modes}
\end{equation}
Then
\begin{equation}
\{\hat d_\mu,\hat d_\nu\}
=\{\hat d_\mu',\hat d_\nu'\}=0,
\qquad
\{\hat d_\mu,\hat d_\nu'\}=\delta_{\mu\nu},
\label{eq:appA_d_CAR}
\end{equation}
and Eq.~\eqref{eq:appA_XY_form} becomes
\begin{equation}
\widehat{\mathcal L}_p
=\sum_{\mu=1}^{2N_s}\beta_\mu\hat d_\mu'\hat d_\mu
+\frac12\hat{\bm d}^{T}\widetilde{\mathbb Y}_p\hat{\bm d},
\qquad
\widetilde{\mathbb Y}_p=R^T\mathbb Y_pR.
\label{eq:appA_L_d_basis}
\end{equation}
The remaining term contains only annihilation maps. Introduce
\begin{equation}
\hat{\bm g}=\hat{\bm d},
\qquad
\hat{\bm g}'=\hat{\bm d}'+C_p\hat{\bm d},
\label{eq:appA_g_shift}
\end{equation}
where the antisymmetric matrix \(C_p\) solves
\begin{equation}
C_p^TB-B^TC_p=\widetilde{\mathbb Y}_p.
\label{eq:appA_C_equation}
\end{equation}
For diagonal \(B\) and \(\beta_\mu+\beta_\nu\neq0\), one may choose
\begin{equation}
(C_p)_{\mu\nu}
=-\frac{(\widetilde{\mathbb Y}_p)_{\mu\nu}}
{\beta_\mu+\beta_\nu},
\qquad
(C_p)_{\nu\mu}=-(C_p)_{\mu\nu}.
\label{eq:appA_C_solution}
\end{equation}
Substitution of Eq.~\eqref{eq:appA_g_shift} into
Eq.~\eqref{eq:appA_L_d_basis} gives
\begin{align}
\widehat{\mathcal L}_p
={}&\sum_\mu\beta_\mu\hat g_\mu'\hat g_\mu
\nonumber\\
&+\frac12\hat{\bm g}^{T}
\left(\widetilde{\mathbb Y}_p-C_p^TB+B^TC_p\right)
\hat{\bm g},
\label{eq:appA_shift_explicit}
\end{align}
so Eq.~\eqref{eq:appA_C_equation} cancels the second line exactly. The normal
form is therefore
\begin{equation}
\widehat{\mathcal L}_p
=\sum_{\mu=1}^{2N_s}\beta_\mu\hat g_\mu'\hat g_\mu.
\label{eq:appA_normal_form_all}
\end{equation}
Because \(C_p\) is antisymmetric, the transformed maps retain the canonical
anticommutation relations.

In the even sector, the right steady state is the normal-mode vacuum when no
additional zero rapidity is present,
\begin{equation}
\hat g_\mu|\rho_{\rm ss}\rangle\!\rangle=0.
\label{eq:appA_right_vacuum}
\end{equation}
The trace functional is defined explicitly by
\begin{equation}
\langle\!\langle\mathbb{I}|A\rangle\!\rangle
\equiv\operatorname{Tr}A,
\label{eq:appA_trace_functional}
\end{equation}
and trace preservation is the unambiguous left-zero-mode condition
\begin{equation}
\langle\!\langle\mathbb{I}|\widehat{\mathcal L}=0.
\label{eq:appA_trace_vacuum}
\end{equation}
The left trace functional and the right steady-state vector are distinct
objects. If an exact single-particle dark mode has zero overlap with the lossy
site, additional zero rapidities and a nontrivial stationary manifold may
occur; Eq.~\eqref{eq:appA_right_vacuum} then does not imply uniqueness.

\subsection{Parity-resolved and full Liouvillian spectra}

Each normal master mode is fermionic. In sector \(p\), the allowed occupation
patterns satisfy
\begin{equation}
\Lambda_{\{n_\mu\}}
=\sum_{\mu=1}^{2N_s}n_\mu\beta_\mu,
\qquad n_\mu\in\{0,1\},
\qquad (-1)^{\sum_\mu n_\mu}=p.
\label{eq:appA_subset_parity}
\end{equation}
There are \(2^{2N_s-1}\) such sums in each parity sector. Their union contains
\(2^{2N_s}=4^{N_s}\) eigenvalues, exactly the dimension of the complete
operator space. The empty subset belongs to \(\mathcal K_+\) and yields the
trace-preserving zero eigenvalue.

For stable dynamics, \(-\Re\beta_\mu\ge0\). The standard full-space gap used in
Eq.~\eqref{eq:Liouvillian_gap_def} is
\begin{equation}
\Delta_{\mathcal L}
=\min_{\mu:\,-\Re\beta_\mu>0}[-\Re\beta_\mu].
\label{eq:appA_gap_beta}
\end{equation}
A parity-even density matrix and a parity-even observable access only even
subset sums. Their sector-resolved edge is
\begin{equation}
\Delta_+
=\min_{\substack{\{n_\mu\}:\,\sum n_\mu\ \text{even}\\
\Lambda_{\{n_\mu\}}\notin\text{stationary manifold}}}
[-\Re\Lambda_{\{n_\mu\}}].
\label{eq:appA_even_gap}
\end{equation}
For the generic geometries used in the intermediate-loss dynamical panels,
there is no exact zero rapidity and the two one-body blocks have identical
lower edges. The smallest even sum then contains one rapidity from each block,
so
\begin{equation}
\Delta_+=2\Delta_{\mathcal L}.
\label{eq:appA_even_gap_twice}
\end{equation}
If exact dark rapidities exist, Eq.~\eqref{eq:appA_even_gap} rather than the
simplified Eq.~\eqref{eq:appA_even_gap_twice} must be used.

\subsection{Independent triangular proof from the no-jump Hamiltonian}

The subset-sum spectrum can also be proved without the superfermion
transformation. Assume first that \(h_{\rm eff}\) is diagonalizable and let
\begin{equation}
\begin{aligned}
h_{\rm eff}|r_\mu\rangle&=\varepsilon_\mu|r_\mu\rangle,\\
\langle l_\mu|h_{\rm eff}&=\varepsilon_\mu\langle l_\mu|,\\
\langle l_\mu|r_\nu\rangle&=\delta_{\mu\nu}.
\end{aligned}
\label{eq:appA_biorth_heff}
\end{equation}
Many-body right eigenstates are biorthogonal Slater states built from the
single-particle modes. Let \(K\) and \(B\) denote occupied mode sets on the ket
and bra sides. The operator
\(|R_K\rangle\langle R_B|\) is an eigenoperator of the no-jump part of
Eq.~\eqref{eq:appA_nojump_form} with eigenvalue
\begin{equation}
\Lambda_{K,B}^{(0)}
=-i\sum_{\mu\in K}\varepsilon_\mu
+i\sum_{\nu\in B}\varepsilon_\nu^*.
\label{eq:appA_nojump_eigs}
\end{equation}
The recycling map \(c_s(\cdot)c_s^\dagger\) lowers both \(|K|\) and \(|B|\)
by one. Ordered by decreasing \(|K|+|B|\), the complete Liouvillian is
therefore lower triangular, and its diagonal entries remain
Eq.~\eqref{eq:appA_nojump_eigs}. Since
\begin{equation}
\operatorname{spec}(M)=\{-i\varepsilon_\mu\},
\qquad
\operatorname{spec}(M^*)=\{i\varepsilon_\mu^*\},
\label{eq:appA_M_heff_spec}
\end{equation}
this independent construction gives exactly the rapidities and parity-resolved
subset sums derived above.

\subsection{Mapping to the plotted matrix \texorpdfstring{$P$}{P}}

For the real SSH hopping matrix,
\begin{equation}
M=-ih-\frac{\Gamma}{2},
\qquad
M^*=ih-\frac{\Gamma}{2}.
\label{eq:appA_M_real}
\end{equation}
Define
\begin{equation}
P_-=-h+\frac{i\gamma}{2}\Pi_s,
\qquad
P_+=h+\frac{i\gamma}{2}\Pi_s.
\label{eq:appA_Ppm_def}
\end{equation}
Then
\begin{equation}
M=iP_-,
\qquad
M^*=iP_+.
\label{eq:appA_M_Ppm}
\end{equation}
If \(P_\pm|u_\mu^{(\pm)}\rangle=E_\mu^{(\pm)}|u_\mu^{(\pm)}\rangle\),
then
\begin{equation}
\beta_\mu^{(\pm)}=iE_\mu^{(\pm)},
\qquad
-\Re\beta_\mu^{(\pm)}=\Im E_\mu^{(\pm)}.
\label{eq:appA_beta_E_pm}
\end{equation}
The SSH chiral matrix
\begin{equation}
\Sigma=\operatorname{diag}(1,-1,1,-1,\ldots,1,-1)
\label{eq:appA_chiral_matrix}
\end{equation}
satisfies \(\Sigma h\Sigma=-h\) and
\(\Sigma\Pi_s\Sigma=\Pi_s\). Therefore
\begin{equation}
\Sigma P_+\Sigma=P_-,
\label{eq:appA_P_similarity}
\end{equation}
so the two blocks have identical spectra. The complete structure matrix nevertheless contains both blocks: each eigenvalue of $P_+$ occurs once in the $P_+$ block and once in the $P_-$ block. Thus the full rapidity multiset contains $2N_s$ entries, while the plotted $N_s\times N_s$ matrix contains all distinct rapidity values. Fermionic subset sums of the duplicated multiset, not of a single $N_s$-element copy, produce the full $4^{N_s}$ Liouvillian spectrum. It is sufficient to plot
\begin{equation}
P\equiv P_+=P_0+\frac{i\gamma}{2}\Pi_s=h_{\rm eff}^*.
\label{eq:appA_P_plotted}
\end{equation}
For a normalized right eigenvector \(P|u\rangle=E|u\rangle\),
\begin{equation}
\Im E=\frac{\gamma}{2}
\frac{\langle u|\Pi_s|u\rangle}{\langle u|u\rangle}\ge0,
\label{eq:appA_ImE_positive}
\end{equation}
which makes the decay-rate convention explicit.

\subsection{Single-mode check and parity sectors}

For \(H=0\) and \(L=\sqrt\gamma\,c\), use the ordered operator basis
\begin{equation}
B_1=|0\rangle\langle0|,\quad
B_2=|0\rangle\langle1|,\quad
B_3=|1\rangle\langle0|,\quad
B_4=|1\rangle\langle1|.
\label{eq:appA_single_basis}
\end{equation}
The Liouvillian actions are
\begin{align}
\mathcal L B_1&=0,
&\mathcal L B_2&=-\frac\gamma2B_2,
\nonumber\\
\mathcal L B_3&=-\frac\gamma2B_3,
&\mathcal L B_4&=\gamma B_1-\gamma B_4.
\label{eq:appA_single_actions}
\end{align}
Thus
\begin{equation}
\mathcal L_{\rm 1m}=
\begin{pmatrix}
0&0&0&\gamma\\
0&-\gamma/2&0&0\\
0&0&-\gamma/2&0\\
0&0&0&-\gamma
\end{pmatrix},
\label{eq:appA_single_matrix}
\end{equation}
with spectrum \(0,\; -\gamma/2,\; -\gamma/2,\; -\gamma\). Here \(B_1,B_4\) are parity even and \(B_2,B_3\) are parity odd. Hence the
full-space gap is \(\gamma/2\), whereas the nonzero even-sector decay edge is
\(\gamma\). The rapidity matrices are \(P_+=P_-=i\gamma/2\), so the two
rapidities are both \(-\gamma/2\). Their even and odd subset sums reproduce the
two parity sectors separately. Replacing \(i\gamma/2\) by \(i\gamma\) would
fail this exact test.

\subsection{Elementwise and spectral numerical validations}

The operator-space formulas were checked with column vectorization,
\(\operatorname{vec}(AXB)=(B^T\otimes A)\operatorname{vec}(X)\). For a
two-site SSH dimer with \(t_1=0.73\), \(\gamma=0.8\), and loss on the first
site, the direct Liouvillian matrix was compared element by element with
Eq.~\eqref{eq:appA_Lp_ab}. The maximum elementwise discrepancy is
\(1.1\times10^{-16}\) in both parity sectors, while the normal-mode
reconstruction from Eq.~\eqref{eq:appA_normal_form_all} gives
\(5.7\times10^{-16}\). For this dimer the full direct spectrum and all
rapidity subset sums agree to \(1.1\times10^{-15}\). The previously used
\(N_{\rm cell}=3\) check (\(N_s=6\), Liouville dimension \(4096\)) gives a
maximum spectral mismatch of \(1.1\times10^{-13}\). These tests verify the
parity sign, the \(\gamma/2\) factor, the structure matrices, the normal-mode
transformation, and the full subset-sum spectrum independently.

For the exact lower-edge EP and the matched control used in
Fig.~\ref{fig:intermediate_gamma}, direct diagonalization of
\(P=P_0+i(\gamma/2)\Pi_s\) gives
\begin{equation}
	\Delta_C=2\Delta_{\mathcal L},\qquad
	\Delta_{\mathcal L}^{\rm EP}\simeq
	\Delta_{\mathcal L}^{\rm ctrl}\simeq1.002824\times10^{-3},
\end{equation}
with \(\Delta_C^{\rm EP}\simeq2.005648\times10^{-3}\) and
\(\Delta_C^{\rm ctrl}\simeq2.005648\times10^{-3}\). This explicitly
confirms that the numerical rates used to compensate the parity-even
correlation dynamics are \(\Delta_C=2\Delta_{\mathcal L}\), not the
parity-odd full-space gap itself.

\subsection{Exceptional points and Jordan form}

The construction of ordinary normal master modes assumes diagonalizability.
At an exact exceptional point, the eigenvalue set is still fixed by the same
structure matrices, but generalized eigenoperators and Jordan chains replace a
complete eigenbasis. Time evolution then contains polynomial factors
\begin{equation}
t^q e^{\Lambda t},
\label{eq:appA_EP_time}
\end{equation}
with \(q\) determined by the Jordan-block size. For a second-order
Jordan block in the one-body drift matrix, a single amplitude has at most a
linear prefactor, whereas a one-body bilinear such as
\(C(t)=e^{Mt}C(0)e^{M^\dagger t}\) can contain polynomial factors up to
second order after the two propagators are combined. This is the sense in which
Fig.~\ref{fig:intermediate_gamma}(d) diagnoses a polynomially enhanced
late-time prefactor. The definitions of the full-space and parity-resolved
decay edges remain spectral definitions based on the real parts of nonzero
eigenvalues; only the eigenoperator expansion must be replaced by its Jordan
form.

\section{Exact one-body dynamics from the many-body Lindblad equation}

\label{app:onebody_dynamics}

In this Appendix we derive the exact real-space one-body dynamics used in
the intermediate-loss figures. The Hamiltonian is number conserving, the
jump operator is linear, and no anomalous pairing terms occur. Although the
physical particle number decreases because of the loss, this passive Gaussian
structure makes the one-body correlation matrix obey an exact closed equation
of motion without any projection onto a reduced few-mode subspace. This exact closure applies not only to local
occupations $n_{j}(t)$ but also to the window observables and compensated
traces used in Sec.~\ref{sec:intermediate_gamma}. We start from the Lindblad equation 
\begin{equation}
\partial _{t}\rho =-i[H,\rho ]+L\rho L^{\dagger }-\frac{1}{2}\{L^{\dagger
}L,\rho \},  \label{eq:app_lindblad}
\end{equation}%
with the quadratic Hamiltonian 
\begin{equation}
H=\sum_{jk}h_{jk}c_{j}^{\dagger }c_{k},  \label{eq:app_h}
\end{equation}%
and the single local loss 
\begin{equation}
L=\sqrt{\gamma }\,c_{s}.  \label{eq:app_L}
\end{equation}%
We define the one-body correlation matrix 
\begin{equation}
C_{jk}(t)\equiv \mathrm{Tr}\!\left[ \rho (t)c_{k}^{\dagger }c_{j}\right] .
\label{eq:app_C_def}
\end{equation}%
The equation of motion is derived most conveniently in the Heisenberg
picture. For an arbitrary operator $O$, 
\begin{equation}
\frac{d}{dt}\langle O\rangle = \bigl\langle i[H,O] + L^{\dagger} O L - \tfrac{1}{2}\{L^{\dagger} L, O\} \bigr\rangle .  \label{eq:app_heis_general}
\end{equation}%
Choosing $O=c_{k}^{\dagger }c_{j}$, we first evaluate the Hamiltonian
contribution. Using 
\begin{equation}
\left[H,c_{j}\right]=-\sum_{\ell }h_{j\ell }c_{\ell },\qquad
\left[H,c_{k}^{\dagger }\right]=\sum_{\ell }h_{\ell k}c_{\ell }^{\dagger },
\label{eq:app_commutators}
\end{equation}%
we obtain 
\begin{equation}
i[H,c_{k}^{\dagger }c_{j}]=-i\sum_{\ell }h_{j\ell }c_{k}^{\dagger }c_{\ell
}+i\sum_{\ell }h_{\ell k}c_{\ell }^{\dagger }c_{j}.
\label{eq:app_H_operator}
\end{equation}%
Taking the expectation value yields 
\begin{equation}
\left. \frac{d}{dt}C_{jk}\right\vert _{H}=-i\sum_{\ell }h_{j\ell }C_{\ell
k}+i\sum_{\ell }C_{j\ell }h_{\ell k},  \label{eq:app_H_C}
\end{equation}%
or, in matrix form, 
\begin{equation}
\left. \dot{C}\right\vert _{H}=-i[h,C].  \label{eq:app_H_matrix}
\end{equation}%
We next evaluate the dissipative contribution. Since 
\begin{equation}
L^{\dagger }L=\gamma c_{s}^{\dagger }c_{s},  \label{eq:app_LdagL}
\end{equation}%
one finds 
\begin{equation}
\left. \frac{d}{dt}C_{jk}\right\vert _{\mathrm{dis}}=-\frac{\gamma }{2}%
\delta _{js}C_{sk}-\frac{\gamma }{2}\delta _{ks}C_{js}.
\label{eq:app_dis_component}
\end{equation}%
Introducing the single-particle loss matrix 
\begin{equation}
\Gamma \equiv \gamma |s\rangle \langle s|,\qquad \Gamma _{jk}=\gamma
\,\delta _{js}\delta _{ks},  \label{eq:app_Gamma}
\end{equation}%
this becomes 
\begin{equation}
\left. \dot{C}\right\vert _{\mathrm{dis}}=-\frac{1}{2}\{\Gamma ,C\}.
\label{eq:app_dis_matrix}
\end{equation}%
Combining Eqs.~\eqref{eq:app_H_matrix} and \eqref{eq:app_dis_matrix}, we
arrive at the exact closed equation 
\begin{equation}
\dot{C}=-i[h,C]-\frac{1}{2}\{\Gamma ,C\}.  \label{eq:app_C_closed}
\end{equation}%
Equivalently, 
\begin{equation}
\dot{C}=MC+CM^{\dagger },\qquad M\equiv -ih-\frac{\Gamma }{2}.
\label{eq:app_MC}
\end{equation}%
The exact solution is therefore 
\begin{equation}
C(t)=U(t)C(0)U^{\dagger }(t),\qquad U(t)=e^{Mt}.  \label{eq:app_C_solution}
\end{equation}%
For a single-particle initial state localized at site $j_{0}$, 
\begin{equation}
|\psi (0)\rangle =c_{j_{0}}^{\dagger }|0\rangle ,\qquad C(0)=|j_{0}\rangle
\langle j_{0}|,  \label{eq:app_single_init}
\end{equation}%
and the local occupation is simply the diagonal part of $C(t)$, 
\begin{equation}
n_{j}(t)=C_{jj}(t).  \label{eq:app_nj}
\end{equation}%
In unit-cell notation this becomes 
\begin{equation}
n_{n,A}(t)=C_{2n-1,2n-1}(t),\qquad n_{n,B}(t)=C_{2n,2n}(t).
\label{eq:app_cell_occ}
\end{equation}%
For the intermediate-loss comparison in the main text, it is convenient to introduce the projector onto the left window,
\begin{equation}
P_L\equiv \sum_{n=1}^{m}\Bigl(|n,A\rangle\langle n,A|+|n,B\rangle\langle n,B|\Bigr).
\label{eq:app_PL_def}
\end{equation}
The corresponding real-space observable is
\begin{equation}
W_L(t)=\mathrm{Tr}[P_L C(t)]
=\sum_{n=1}^{m}\bigl[n_{n,A}(t)+n_{n,B}(t)\bigr],
\label{eq:app_WL_def}
\end{equation}
which is the quantity plotted in Fig.~\ref{fig:intermediate_gamma}(c). Because $C(t)$ is parity even, its exponential rates are sums $\beta_\mu^{(-)}+\beta_\nu^{(+)}$. At the two parameter points used in the comparison, the slow one-body edge is $\Delta_C=2\Delta_{\mathcal L}$. The compensated signal shown in Fig.~\ref{fig:intermediate_gamma}(d) is therefore
\begin{equation}
\widetilde W_L(t)=
\frac{e^{\Delta_C t}W_L(t)}{e^{\Delta_C t_*}W_L(t_*)},
\label{eq:app_WL_comp}
\end{equation}
where $t_*$ is the left end of the displayed late-time window.

Equations~\eqref{eq:app_C_closed}--\eqref{eq:app_WL_comp} are the exact
real-space one-body dynamics used for the numerical time traces in the main
text. They show that the plotted local occupations, window occupations, and compensated traces are exact observables of the
original many-body Lindblad problem for the chosen initial states, rather
than a projected few-mode approximation. As explained in Sec.~\ref{subsec:statistics_experiment}, the same drift matrix governs one-body correlations in linear bosonic systems and field amplitudes in classical-wave implementations, although these platforms do not possess the finite fermionic parity-resolved spectrum derived in Appendix~\ref{app:third_quantization}.
\begin{figure*}[!t]
\centering
\includegraphics[width=\textwidth]{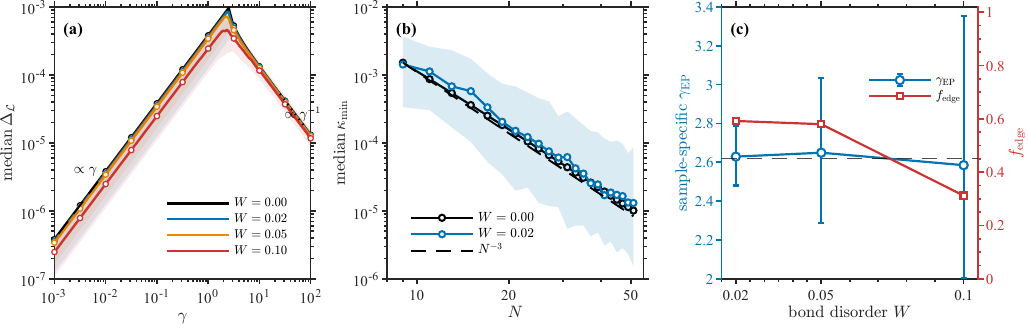}%
\caption{Robustness against chiral-symmetry-preserving bond disorder.
(a) Ensemble median of the full-space Liouvillian gap for $N=15$,
$t_1=0.655$, $t_2=1$, and loss on $(m,A)=(8,A)$. For each nonzero $W$,
the shaded region is the $25\%$--$75\%$ realization interval from $200$
samples. The weak- and strong-loss guides show
$\Delta_{\mathcal L}\propto\gamma$ and
$\Delta_{\mathcal L}\propto\gamma^{-1}$, respectively.
(b) Median weak-loss coefficient
$\kappa_{\min}=\min_{\alpha:\,|\phi_\alpha(s)|^2>0}|\phi_\alpha(s)|^2$
for a trivial chain with $t_1=1$, $t_2=0.6$, central loss, and odd
$N=9,11,\ldots,51$. The finite-size statistics use $500$ realizations for
each displayed nonzero disorder strength; the dashed line is an $N^{-3}$ guide.
(c) For the geometry of panel (a), the left axis shows the median and
$16\%$--$84\%$ interval of the exact EP with the smallest positive
$\operatorname{Im}E$ in each realization. The dashed horizontal line is the
clean value $\gamma_{\mathrm{EP}}=2.62$. The right axis shows
$f_{\mathrm{edge}}$, the fraction of successfully identified exact EPs that
also satisfy
$\operatorname{Im}E_{\mathrm{EP}}/\Delta_{\mathcal L}\leq1.01$.
The exact-EP search succeeds for all $500$ realizations at each displayed
disorder strength.}
\label{fig:disorder_robustness}
\end{figure*}

\section{Robustness against weak bond disorder}
\label{app:disorder}

To test the stability of the three-regime picture, we introduce
chiral-symmetry-preserving bond disorder,
\begin{equation}
t_{1,n}=t_1\left(1+W\xi_{1,n}\right),\qquad
t_{2,n}=t_2\left(1+W\xi_{2,n}\right),
\label{eq:disorder_bonds}
\end{equation}
where the independent random variables $\xi_{1,n}$ and $\xi_{2,n}$ are
uniformly distributed in $[-1,1]$. No onsite disorder is added, so the
disordered hopping matrix $h_W$ remains bipartite and obeys
$\Sigma h_W\Sigma=-h_W$. For every realization we use
$P_W=h_W+i(\gamma/2)\Pi_s$, with the same physical-loss convention as in the
main text. The calculation therefore tests coupling inhomogeneity without
explicitly breaking the chiral symmetry that organizes the clean SSH
spectrum.

The weak-loss relation remains valid realization by realization:
$\operatorname{Im}E_\alpha^{(W)}
=(\gamma/2)|\phi_\alpha^{(W)}(s)|^2+O(\gamma^2)$.
Fig.~\ref{fig:disorder_robustness}(a) confirms that the ensemble-median gap
has a weak-loss exponent indistinguishable from unity for all displayed
values of $W$. In Fig.~\ref{fig:disorder_robustness}(b), fits over the
displayed finite-size range give exponents $-2.926$ for the clean chain and
$-3.050$ for $W=0.02$. Thus the clean $N^{-3}$ trend remains visible for
genuinely weak coupling disorder. We do not infer an asymptotic disordered
power law from stronger $W$, where one-dimensional localization corrections
can become important.

The strong-loss mechanism is equally insensitive to bond randomness at the
level of its exponent. Eliminating the lossy site in each realization gives
$P_{Z,W}=P_{R,W}+(2i/\gamma)\bm v_W\bm v_W^T+O(\gamma^{-2})$.
Disorder changes the realization-dependent geometric prefactor, but not the
$\gamma^{-1}$ scaling. Fits to Fig.~\ref{fig:disorder_robustness}(a) yield
strong-loss exponents between $-1.0021$ and $-1.0002$ for
$0\leq W\leq0.10$.

The intermediate regime is more sensitive. Because
Eq.~\eqref{eq:disorder_bonds} preserves chiral symmetry, exact imaginary-axis
EPs remain possible sample by sample. Direct diagonalization at the
numerically located roots gives median pair distances of approximately
$1.5\times10^{-9}$ and median phase rigidities of approximately
$7.5\times10^{-7}$ for all three nonzero disorder strengths. Their positions,
however, broaden: the median values of $\gamma_{\mathrm{EP}}$ are $2.629$,
$2.649$, and $2.585$ for $W=0.02$, $0.05$, and $0.10$, respectively, with the
corresponding $16\%$--$84\%$ intervals shown in
Fig.~\ref{fig:disorder_robustness}(c). We define \(N_{\mathrm{EP}}\) as the
number of successful exact-EP searches and \(N_{\mathrm{edge}}\) as the subset
that also satisfies
\(\operatorname{Im}E_{\mathrm{EP}}/\Delta_{\mathcal L}\leq1.01\). The reported
fraction is
\begin{equation}
	f_{\mathrm{edge}}=\frac{N_{\mathrm{edge}}}{N_{\mathrm{EP}}}.
\end{equation}
It decreases from $0.592$ and $0.580$ at $W=0.02$ and $0.05$ to $0.312$ at
$W=0.10$. The selected object is the exact EP with the smallest positive
imaginary energy in a given realization; under disorder it need not be the
continuous descendant of one fixed clean pair. Bond disorder therefore
preserves exact EPs but shifts and splits the simultaneous clean-system
coalescences, so an exact EP does not necessarily remain the
gap-controlling lower-edge singularity.

These results establish a qualified robustness statement. The local-weight
selection rule and the Zeno $\gamma^{-1}$ law are robust organizing
mechanisms, and the clean $N^{-3}$ finite-size behavior survives genuinely
weak bond disorder over the sizes considered. By contrast, the location and
lower-edge dominance of the intermediate exact EP are realization dependent.
Onsite disorder, which breaks chiral symmetry, would generically relax the
balanced-weight condition and unfold an exact EP on the real-$\gamma$ axis into near-EP
behavior; that distinct symmetry-breaking problem is not included in the
present robustness test.

\end{document}